\documentclass[fleqn,usenatbib]{mnras}

\usepackage{newtxtext,newtxmath}

\usepackage[T1]{fontenc}
\usepackage{rotating}
\makeatletter
\let\@makecaption=\SFB@makefigurecaption
\makeatother
\usepackage{tabularx}

\DeclareRobustCommand{\VAN}[3]{#2}
\let\VANthebibliography\thebibliography
\def\thebibliography{\DeclareRobustCommand{\VAN}[3]{##3}\VANthebibliography}

\usepackage{graphicx}	
\usepackage{amsmath}	
\usepackage{float}
\usepackage{bm}
\usepackage{booktabs}
\usepackage{orcidlink}
\hypersetup{colorlinks=true, citecolor=blue, linkcolor=blue, urlcolor=blue}

\title[Ubiquitous N-Enrichment in $6 \leq z \leq 10$ Galaxies]
{Unveiling and Characterising Ubiquitous Nitrogen Enhancement in $\bm{6 \leq z \leq 10}$ Galaxies with JWST Spectroscopy}

\author[Rai \& Roberts-Borsani]{
Raunaq Singh Rai\,\orcidlink{0009-0000-8489-7211}$^{1}$\thanks{E-mail: raunaq.rai.25@ucl.ac.uk}
and Guido Roberts-Borsani\,\orcidlink{0000-0002-4140-1367}$^{1}$
\\
$^{1}$Department of Physics \& Astronomy, University College London, London, WC1E 6BT, UK\\
}

\date{Accepted XXX. Received YYY; in original form ZZZ}

\pubyear{\the\year{}}

\begin{document}
\label{firstpage}
\pagerange{\pageref{firstpage}--\pageref{lastpage}}
\maketitle

\begin{abstract}
The James Webb Space Telescope (JWST) has revealed a growing number of $z>6$ galaxies with enhanced N/O ratios that challenge standard chemical enrichment frameworks. We test whether this is exclusive to extreme systems or a generic feature of $6 \leq z \leq 10$ populations through a stacking analysis of 135 galaxies with JWST/NIRSpec $R\sim1000$ spectroscopy, probing the origin of N, C and O enrichment and its dependence on recent star formation. A full-sample composite reveals prominent C III], C IV, N IV] and He II emission, demonstrating a ubiquity of high-ionisation UV lines. Auroral [O III]$\lambda$4363 electron temperatures and multi-ionisation zone modelling yield direct-$T_e$ abundances with supersolar $\log(\mathrm{N/O}) = -0.57^{+0.09}_{-0.10}$ and subsolar $\log(\mathrm{C/O}) = -0.80^{+0.03}_{-0.03}$ at $12 + \log(\mathrm{O/H}) = 7.79^{+0.03}_{-0.03}$, in contrast with local samples and suggesting significant nitrogen enhancement within the first billion years. Stacking further by recent star formation activity over burst timescales of $\Delta t = 3$--$20$~Myr, we find that sources caught in a recent burst are less enriched in both N/O and metallicity than those in a relative lull, with the largest contrasts confined to the shortest timescales. C/O, however, remains largely invariant across all metrics. We interpret these observations as the result of delayed AGB enrichment across multiple burst episodes, together with dilution by pristine gas inflows. Cosmological simulations reproduce this trend with CCSNe and AGB yields alone, without the need for shorter-lived (super-)massive stars. Enhanced N/O therefore appears to be a generic feature of high-redshift systems, produced by standard enrichment processes rather than requiring new physical frameworks.

\end{abstract}

\begin{keywords}
galaxies: abundances -- galaxies: high-redshift -- galaxies: ISM -- galaxies: evolution
\end{keywords}



\section{Introduction}

The first billion years of cosmic history (corresponding to redshifts $z>6$) represent a crucial time period in which the first stars, galaxies and black holes formed, the (re)ionisation of the neutral intergalactic medium (IGM) took place, and the first heavy elements were synthesised to yield the complex chemistry we see today. As such, pinpointing and characterising the formation and evolution of sources within this epoch represents a fundamental goal of extragalactic astrophysics. To this end, the metallicities, elemental abundance ratios, and chemical histories of the most distant sources serve as key diagnostics for past star formation and early galaxy evolution \citep[e.g.,][]{maiolino2019, kobayashi2020}. Heavy elements such as carbon, nitrogen, and oxygen are synthesised through successive episodes of star formation and released into the interstellar medium (ISM) through supernovae and stellar winds on distinct timescales: oxygen and carbon are returned within a few $\sim$10 Myr predominantly by core-collapse supernovae (CCSNe), while nitrogen is returned either within a few Myr by Wolf Rayet and (super)massive star winds or over $\sim$100--300 Myr by asymptotic giant branch (AGB) stars (e.g., \citealt{nomoto2013, karakas2010, berg2025}). The accumulation of these metals and their abundance ratios therefore encode specific stellar enrichment pathways, the influence of gas inflows and outflows, and the possible signatures of Population III and II stellar generations.

The advent of the James Webb Space Telescope (JWST) has now enabled precision measurements of the elemental abundances, ionising conditions, and star formation histories of $z>6$ sources, through direct spectroscopic measurements. These have since been traced back to the first few hundred Myr after the Big Bang, and as far out as $z = 14.4$ \citep{naidu2026}, revealing rapidly enriched systems and extreme ionising conditions that differ significantly to those seen in the local Universe \citep[e.g.,][]{curtislake2023, arrabal2023, carniani2024, castellano2024, napolitano2025, harikane2024, robertsborsani2024, robertsborsani2026, hsiao24, zavala25, alvarezmarquez25, alvarezmarquez26, marqueschaves26}. A major surprise in some of these JWST observations has been the revelation of high-ionisation UV nitrogen lines (N\,\textsc{iv}]\,$\lambda\lambda$1483,1486 \AA\ and/or N\,\textsc{iii}]\,$\lambda\lambda$1749,1752 \AA) and supersolar N/O ratios in a number of especially luminous ($M_{\rm UV}<-20$) $z>10$ sources -- e.g., GHZ9 at $z=10.1$ \citep{napolitano2025}, GN-z11 at $z=10.6$ \citep{bunker2023gnz11},  GHZ2 at $z=12.3$ \citep{castellano2024}, and MoM-z14 at $z=14.4$ \citep{naidu2026}, amongst others -- indicating significant nitrogen-enrichment within only a few hundred Myr of evolutionary time. 

Since then, nitrogen enhancement has also been revealed in a growing number of extreme sources down to $z\sim5$ (e.g., \citealt{ji2024, marques2024, topping2024, isobe2023, toppingCN, schaerer2024, berg2025}), suggesting a higher prevalence than previously thought and a possible generic origin. Such anomalous abundance ratios, which contrast with local samples of comparable metallicity \citep{berg2019, Ji2025}, are challenging to reconcile with standard chemical frameworks and theoretical expectations given the traditionally long timescales associated with substantial enrichment by AGB stars. Alternative, faster and more exotic enrichment channels have also been proposed to reconcile these timescales, such as tidal disruption events around black holes \citep{cameron2023,watanabe2024,ji2024,isobe25}, contributions from Wolf Rayet stellar wind ejecta and globular cluster precursors \citep{senchyna2024,watanabe2024,isobe2023,nagele2023,berg2025,naidu2026,schaerer2026gc}, and even very-to-super massive stars (VMS and SMS, $M_{*}>10^{2-3} M_{\odot}$; \citealt{charbonnel2023,marques2024,marqueschaves2026}). Indeed, targeted JWST/NIRSpec observations have revealed such contributions in only a few marked examples: \citet{berg2025} presented a prominent optical ``blue bump'' from multiple high-ionisation nebular lines (N\,\textsc{iii}$\lambda\lambda$4634,4642 \AA, C\,\textsc{iii}$\lambda\lambda$4649,4667 \AA, Fe\,\textsc{iii}$\lambda$4660 \AA, and He\,\textsc{ii}$\lambda$4687 \AA), together with broadened Balmer, [O\,\textsc{iii}] and He\,\textsc{ii}\,$\lambda\lambda1640,4687$ components, indicative of Wolf Rayet features in the lensed RXCJ2248-ID3 nitrogen-enriched source at $z=6.1$; similarly, \citet{marqueschaves2026} showcase strong P-Cygni profiles in N\,\textsc{v}$\lambda$1240 \AA\ and C\,\textsc{iv}$\lambda\lambda$1548,1550 \AA\ wind lines, suggestive of very massive star signatures in two luminous $z\sim8.7$ sources.

Despite these remarkable observations, the majority of these fast channels rely on stellar populations that are rare, short-lived, or both. Accommodating them as a generic feature of the high-redshift population would require invoking non-standard physics, either as a top-heavy IMF extending well beyond the typical upper mass limit ($M_{\rm upper}\gtrsim 300$--$1000\,M_\odot$) to populate the VMS/SMS regime, or a large fraction of galaxies hosting extreme-density ($\rho\gtrsim 10^6\,M_\odot\,{\rm pc}^{-3}$) cluster cores in which SMS can form via runaway stellar collisions \citep{portegieszwart02,fujii24}. Even for Wolf Rayet stars, the $\sim 0.3$\,Myr nitrogen-enriched wind phase \citep{berg2025} and rarity at low (sub-Small Magellanic Cloud) metallicities \citep{crowther2007} place stringent bounds on how much of the early global galaxy population could plausibly be enriched by these channels at any given epoch.



Considering the vastly different ISM conditions at high redshift, where galaxies are denser and more highly ionised than their local counterparts \citep[e.g.,][]{cameron2023b, isobe2023ne}, and the more intense, burstier star formation that sustains them \citep[e.g.,][]{fauchergiguere2018, reddy2023, pallottini2023, looser2025}, a more generic explanation for anomalous chemical abundances remains plausible. However, population-level constraints have thus far been prevented by important instrumental, sample and physical biases: most N\,\textsc{iv}] detections have to date been confined to extreme and luminous sources, relying on low-resolution ($R \sim 30$--$50$ at UV wavelengths) NIRSpec prism spectroscopy and strong-line derived metallicities, biasing not only the samples themselves but the physical properties inferred from them.
To make progress, and determine whether nitrogen enrichment from high-ionisation UV lines at the earliest times represents a signpost of exotic physics in a handful of extreme sources, or a generic feature from unexplored modes of early star formation, requires representative galaxy samples with precise abundance measurements from high-resolution spectroscopy with sufficient depth.

In this paper we aim to address this question by using a stacking analysis of NIRSpec $R\sim1000$ spectra to determine the prevalence of high-ionisation N, C, and O lines, derive accurate abundances with electron temperature and multi-ionisation zone modelling, and compare their observed ratios to recent star formation histories and cosmological simulations with varying stellar origins.
The paper is organised as follows. The observations and sample selection are described in Section~\ref{sec:data_methods}, followed by the methods used in this work in Section~\ref{sec:methods} (covering the stacking procedure, spectroscopic measurements, spectral energy distribution fitting, and direct-method chemical abundance determinations). We present composite spectra and their characteristic properties in Section~\ref{sec:full_stack} and determine their dependence on recent star formation history in Section~\ref{sec:binned}. Finally, in Section~\ref{sec:discussion} we compare our results with galactic chemical evolution models and cosmological simulations, and discuss the physical processes plausibly driving the elemental abundance patterns we observe.

Throughout this paper we adopt the flat $\Lambda$CDM cosmology of \citet{planck2018}, with $H_{0} = 67.7$~km~s$^{-1}$~Mpc$^{-1}$, $\Omega_{\rm m} = 0.31$ and $\Omega_{\Lambda} = 0.69$. All magnitudes are quoted in the AB system \citep{oke1983}. Elemental abundances are reported on the standard logarithmic scale, with the solar reference values of \citet{asplund2009} ($12 +\log(\mathrm{O}/\mathrm{H})_{\odot} = 8.69$, $\log(\mathrm{N}/\mathrm{O})_{\odot} = -0.86$, $\log(\mathrm{C}/\mathrm{O})_{\odot} = -0.26$, $\log(\mathrm{Ne}/\mathrm{O})_{\odot} = -0.76$). Unless stated otherwise, we use metallicity to refer to the gas-phase oxygen abundance, $12 + \log(\mathrm{O}/\mathrm{H})$.

\section{Data and Sample Selection}
\label{sec:data_methods}

\subsection{JWST/NIRSpec Spectroscopy}

We consider publicly available extragalactic NIRSpec micro-shutter assembly (MSA) spectroscopy obtained with R$\sim1000$ gratings (G140M-F070LP, G140M-F100LP, G235M-F170LP, G395M-F290LP). Where available, we additionally include any overlapping prism spectroscopy ($R \sim 30$--$300$), which is more sensitive to continuum and strong-line measurements.

The data sets are drawn from the JADES (\citealt{jades1,jades2,jades3,jades4}; PI Eisenstein), CEERS (\citealt{ceers1}; PI Finkelstein), EXCELS (\citealt{excels1}; PI Carnall), AURORA (\citealt{aurora}; PI Shapley), GLASS (\citealt{glass,glass2}; PI Treu ), RUBIES (\citealt{rubies}; PI de Graaff), CAPERS (\citealt{capers}; PI Dickinson), UNCOVER (\citealt{uncover1}; PI Labb\'e) and SPURS (\citealt{spurs}; PI Mason and Stark) surveys, and span the GOODS-S, GOODS-N, EGS, UDS, and Abell 2744 fields. NIRCam photometry, useful for flux loss considerations (as described below) and continuum measurements, is taken from the ASTRODEEP catalogue \citep{astrodeep}, which provides a consistent reduction of wide-band measurements across each of the considered fields. Data reduction was performed using the \texttt{msaexp} pipeline (v0.9.17; \citealt{msaexp})\footnote{\url{https://github.com/gbrammer/msaexp}} together with the \texttt{jwst} pipeline (v1.19.2) and the \texttt{jwst\_1464.pmap} CRDS files, following the procedures broadly outlined by \citet{robertsborsani2024} and \citet{meyer25} and described in detail in \citet{rubies} and \citet{primal}. We refer the reader to those papers for a detailed description. 

To correct for potential flux losses due to the small MSA shutter sizes, we applied corrections to the 1D prism spectra. Synthetic photometry was computed by convolving the spectrum with each NIRCam photometric band and deriving flux loss ratios with the true photometry. A wavelength-dependent correction was obtained by fitting a second-order polynomial to these ratios, and applied to the spectrum. No flux loss corrections were applied to the grating spectra, since individual spectra rarely reach the depth necessary for a robust continuum measurement. For each reduced source, we combine all available grating spectra onto a common wavelength grid and combine overlapping regions between adjacent gratings via inverse-variance weighted averaging. 

\subsection{Sample Selection of $6 \leq z \leq 10$ Galaxies}
\label{sec:sample}


Spectroscopic redshifts were determined from the combined grating spectra using the catalogues of \citet{robertsborsani2024, robertsborsani2026} as a prior, and a redshift-fitting pipeline that minimises the $\chi^2$ of a least-squares fit to a set of strong rest-frame optical emission lines (all modelled simultaneously), principally [O\,\textsc{ii}]\,$\lambda\lambda3726,3729$, [Ne\,\textsc{iii}]\,$\lambda3869$, [O\,\textsc{iii}]\,$\lambda\lambda4960,5008$, H$\beta$, and H$\alpha$, each modelled as a Gaussian of fixed instrumental width at its redshifted wavelength, over a grid of possible redshifts.

Given our interest in galaxies within the first billion years of the Universe, we restrict our sample to sources with a secure redshift of $6 \leq z \leq 10$, such that the entire rest-frame UV out to H$\beta$ and [O\,\textsc{iii}]\,$\lambda$5008 is covered for each of our sources. Moreover, given detector gaps present in some individual spectra, we require each source to have $R \sim 1000$ spectroscopic wavelength coverage of N\,\textsc{iv}]\,$\lambda\lambda$1483,1486, C\,\textsc{iv}\,$\lambda\lambda$1548,1550, N\,\textsc{iii}]\,$\lambda\lambda$1749,1752, C\,\textsc{iii}]\,$\lambda\lambda$1907,1909, [O\,\textsc{iii}]\,$\lambda$4363 and [O\,\textsc{iii}]\,$\lambda$5008. We additionally impose a minimum signal-to-noise cut on the rest-frame UV continuum of each spectrum. Known active galactic nuclei (AGN) from the literature \citep[e.g.,][]{jadesagn1,jadesagn2,ceersagn1} were excluded, along with any sources showing broad Balmer components (H$\beta$ or H$\alpha$) without corresponding broad [O\,\textsc{iii}]\,$\lambda$4960 and [O\,\textsc{iii}]\,$\lambda$5008 components, indicative of a Broad Line Region in an AGN.

Where prism spectra are available, we use the median flux density between $1400 < \lambda_{\text{rest}} < 1600$\,\AA\ to calculate the absolute UV magnitude ($M_{\text{UV}}$), otherwise we follow the procedure outlined in \citet{bouwens2015} and interpolate between photometric bands bracketing $\lambda_{\text{obs}} = 1500\,(1+z)$\AA. For sources behind the Abell 2744 galaxy cluster, we correct for lensing magnification using the lens model of \citet{abell2744lense}. An absolute UV magnitude cut of $-21 < M_{\rm UV} < -19$ is applied to our sample in order to minimise any luminosity-dependent correlations within the sample. This range corresponds approximately to $0.1$--$1\,L^{*}$ at $6 \leq z \leq 10$ \citep{schechter1976,bouwens2015}. As a final sanity check, all resulting spectra were visually inspected, and any source affected by e.g. poor background subtraction, residual artefacts, contamination from nearby objects, etc, was removed from the final sample.

\begin{figure}
    \centering
    \includegraphics[width=1\linewidth]{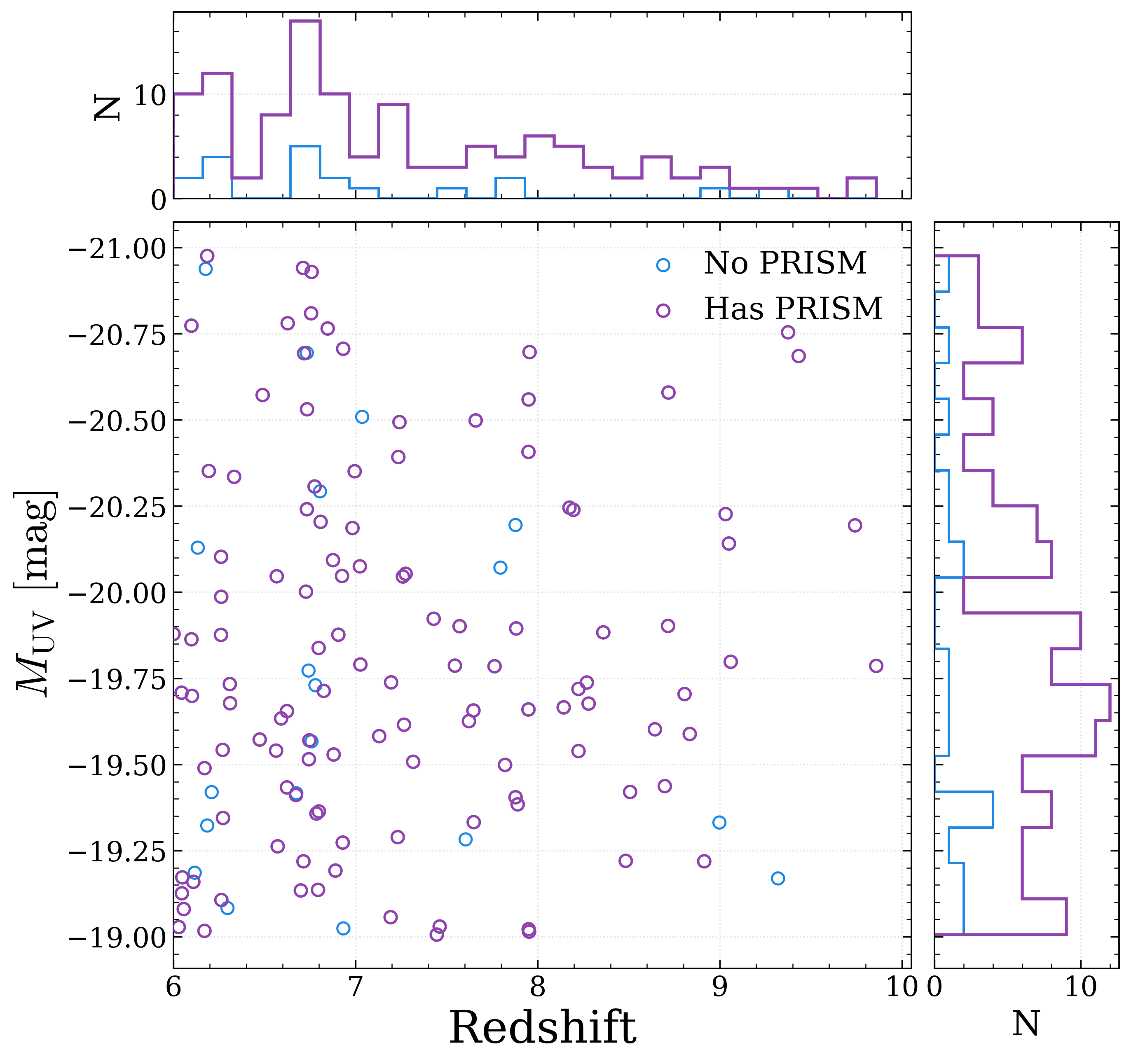}
    \caption{Absolute UV magnitude versus redshift for our fiducial sample of $6 \leq z \leq 10$ galaxies with $R \sim 1000$ spectroscopy, colour-coded by prism availability (blue circles represent sources with grating spectra only, while purples sources have both grating and prism spectra). Histograms show the redshift distribution (top) and $M_{\mathrm{UV}}$ distribution (right).}
    \label{fig:m_uv_redshift}
\end{figure}

This results in a fiducial sample of 135 emission-line-selected galaxies at $6 \leq z \leq 10$ with R$\sim1000$ spectroscopy, 117 of which also harbour prism spectroscopy. The sample is shown in a $M_{\rm UV}$--$z_{\rm spec}$ plane in Figure~\ref{fig:m_uv_redshift}, and its breakdown by field is given in Table~\ref{tab:field_stats}.

\begin{table}
    \centering
    \footnotesize
    \setlength{\tabcolsep}{4pt}
    \renewcommand{\arraystretch}{1.3}
    
    \begin{tabular}{l >{\centering\arraybackslash}p{3.6cm} c c}
    \hline \hline
    \textbf{Field} & \textbf{Survey(s) (PID)} & \textbf{Total} & \textbf{With prism} \\
    \hline
    GOODS-S & JADES (1180, 1210, 3215, 1286, 1287) & 71 & 71 \\
    GOODS-N & SPURS (9214), AURORA (1914), JADES (1181) & 34 & 28 \\
    EGS & SPURS (9214), CEERS (1345), CAPERS (6368), RUBIES (4233) & 23 & 17 \\
    UDS & EXCELS (3543) & 1 & 0 \\
    Abell 2744 & SPURS (9214), GLASS (1324), UNCOVER (2561) & 6 & 1 \\
    \hline
    \textbf{Total} & & \textbf{135} & \textbf{117} \\
    \hline
    \end{tabular}
    \caption{The observed fields and spectroscopic JWST/NIRSpec programmes from which our $6 \leq z \leq 10$ galaxy sample is derived. The ``Total'' column refers to the number of sources with $R \sim 1000$ spectroscopy, while the ``With prism'' refers to the subset of the total also observed with prism spectroscopy.}
    \label{tab:field_stats}
\end{table}

\section{Methods}
\label{sec:methods}

\subsection{Spectral Stacking}
\label{sec:stack_method}

Faint diagnostic emission lines, in particular UV nitrogen lines N~{\sc iv}] and N~{\sc iii}], are rarely detected in individual spectra at current survey depths. \citet{zhu2025} show that such lines are subject to a strong observational bias, where photoionisation models with sub-solar N/O fall below the detection thresholds of current surveys (e.g., CEERS with $\log(\mathrm{N/O}) \gtrsim -0.4$, and JADES with $\log(\mathrm{N/O}) \gtrsim -1.0$) and thus only the very strongest lines are observable. The auroral [O\,{\sc iii}]\,$\lambda$4363 line required for ``direct'' electron temperature-based metallicities (described below) poses a further challenge, since it is not recoverable at the depths and low spectral resolution of prism spectroscopy. As a result, most observations blend the line with adjacent H$\gamma$ emission, and even when resolved it is detected in only a minority of $z > 6$ sources at current depths \citep[e.g.,][]{curti2023, laseter2024, sanders25, pollock2026b}.

We therefore employ spectral stacking, which boosts the effective SNR through a median composite spectrum, and allows us to recover these otherwise inaccessible lines in representative bins of a set of galaxy properties. By co-adding $N$ spectra, uncorrelated noise averages down relative to common spectral features, raising the SNR by a factor of $\sqrt{N}$. Prior to stacking, the grating spectrum of each galaxy is shifted to the rest frame using its spectroscopic redshift, $\lambda_{\rm rest} = \lambda_{\rm obs}/(1+z)$, and interpolated onto a common rest-frame grid spanning $1000$--$7500$~\AA\ at $0.5$~\AA. This grid oversamples the native NIRSpec $R\sim1000$ sampling across our full redshift range, so the resampling does not degrade the narrow diagnostic lines. Each spectrum is then normalised by its median flux in the rest-frame $1700$--$1850$~\AA\ continuum window. This range was chosen as a window of UV continuum free of strong emission lines that lies close to the nitrogen diagnostics of interest and is covered by the grating spectra across our full redshift range. The continuum-normalised spectra are then added to a stack according to a binning of choice, and a high-SNR composite is derived as the median of the fluxes in each wavelength bin. We estimate the uncertainty on the spectrum by bootstrap resampling: in each of $1000$ trials we resample the $N$ contributing spectra with replacement and recompute the median. This estimate captures the galaxy-to-galaxy variance within each bin, which is typically larger than the average noise determined from the $1\sigma$ uncertainty array of the individual spectra.

\subsection{Spectroscopic Measurements}
\label{sec:linefitting}

Our analysis rests on accurate emission-line and continuum measurements. We measure the former with a single fitting procedure, applied identically to the composite and individual spectra so that all measurements are derived self-consistently.

The continuum is estimated from a moving average across the spectrum, computed after masking the regions around known emission lines, and is then subtracted. Each emission line is modelled as a Gaussian which, rather than being evaluated at the pixel centres, is integrated analytically over the wavelength range spanned by each spectral pixel. Every line is fit with an amplitude (the integrated flux), a central wavelength, and a width, with each centroid allowed to shift from its rest-frame position by up to $500\,\mathrm{km\,s^{-1}}$. We fit these parameters by maximising the Gaussian log-likelihood, $\ln\mathcal{L}$:

\begin{equation}
  \ln\mathcal{L} = -\tfrac{1}{2}\sum_{i}\left[\frac{(f_{\lambda,i}-m_{\lambda,i})\,w_{i}}{\sigma_{\lambda,i}}\right]^{2},
  \label{eq:loglike}
\end{equation}
where $f_{\lambda,i}$ and $\sigma_{\lambda,i}$ are the observed flux and its uncertainty in pixel $i$, $m_{\lambda,i}$ is the pixel-integrated model, and $w_{i}$ is a per-pixel weight. We obtain full posteriors via Markov Chain Monte Carlo (MCMC) sampling with the No-U-Turn Sampler \citep[NUTS;][]{nuts} as implemented in \textsc{NumPyro} \citep{phan2019composable}. The entire likelihood is written in \textsc{jax}, which provides just-in-time compilation and automatic differentiation.
We run $8$ independent chains, each with $500$ warm-up (burn-in) iterations that are discarded while the sampler tunes how it steps through parameter space, followed by $2000$ retained samples per chain. To ensure convergence, we require the Gelman--Rubin statistic to satisfy $\hat{R} < 1.01$ for all fitted parameters. We report each quantity, including the measured line fluxes, as the posterior median, with uncertainties given by the 16th and 84th percentiles of the posterior.

\subsection{Spectral Energy Distribution Fitting}
\label{sec:sed_fitting}

Star formation histories are central to this work, given a key aim is to establish how a galaxy's recent star formation regulates its gas-phase chemical abundances and their observed ratios.
In order to determine star formation histories (SFHs), star formation rates (SFRs), and stellar masses ($M_{*}$), useful to characterise our sample of galaxies, we adopt spectral energy distribution (SED) fitting using the Bagpipes \citep{bagpipes1} code. For sources with prism spectra (117 sources), the fit was performed jointly on the spectroscopy and NIRCam photometry; otherwise, photometry alone was used (18 sources).

A non-parametric star formation history \citep{nonparametricsfh} was used with a Student-t prior on the log SFR ratio between adjacent bins, permitting both smooth evolution and bursty episodes \citep{tacchella22}. Nebular emission was modelled using pre-computed grids generated with \texttt{Cloudy} v25.00 \citep{cloudy2025}, based on BPASS
v2.2.1 stellar populations \citep{eldridge2017} with the broken power-law initial mass function (IMF) of \citet{stanway}. For dust attenuation we adopt the \citet{salim} law, which generalises \citet{calzetti2000} with a power-law slope deviation ($\delta$) and UV bump amplitude ($B$). Fixed empirical curves can underestimate intrinsic luminosities at high redshift \citep{koprowski2020}, while high-redshift and low-metallicity galaxies frequently require steeper, SMC-like attenuation \citep[e.g.,][]{reddy2015, shivaei2020, markov2024,reddy2026}. We assume $R_V = 3.15$, with $\delta$ and $B$ left as free parameters with Gaussian priors ($\delta = -0.35 \pm 0.2$; $B = 2.27 \pm 1.0$) centred on a curve slightly steeper than \citet{calzetti2000} with a moderate UV bump, typical of star-forming galaxies, while broad enough to reach the steeper, bump-free SMC-like regime where the data require it. The nebular emission grids are parametrised by the ionisation parameter $\log U$ and the Lyman-continuum escape fraction $f_{\text{esc}}$ \citep{giovinazzo26}. We adopt a broad uniform prior on $\log U$ ($-3.5 < \log U < 0$) and a log-uniform prior on $f_{\text{esc}}$ ($0.001 < f_{\text{esc}} < 0.5$).
We additionally marginalise over a multiplicative spectroscopic calibration term with a Gaussian prior centred on unity ($\mu = 1.0$, $\sigma = 0.2$). Despite the flux loss corrections discussed earlier, this accounts for any remaining artefacts that could offset the 
NIRSpec spectroscopy from that of the NIRCam photometry \citep{bagpipes1}, rather than reconciling potential mismatches through the physical parameters.
The full set of parameters and their priors are listed in Table~\ref{tab:sed_priors}.

\begin{table}
\centering
\resizebox{\columnwidth}{!}{%
    \begin{tabular}{llp{4.5cm}}
    \hline \hline
    \textbf{Parameter} & \textbf{Prior / Fix} & \textbf{Range / Value} \\ \hline
    Redshift ($z$) & Gaussian & $\mu=z_{\text{spec}}, \sigma=0.05$ \\
    $\log_{10}(M_*/M_\odot)$ & Log-Uniform & $5 < \log M_* < 13$ \\
    Metallicity ($Z$) & Log-Uniform & $10^{-3} < Z < 0.5$ \\
    $\log(\text{SFR}_{i+1}/\text{SFR}_i)$ & Student-t & $\nu=2, \sigma=0.3$ \\
    SFH Time Bins (Myr) & Fixed & 3, 5, 10, 20, 50, \newline 100, 200, 500 \\
    Extinction ($A_V$) & Uniform & $0 < A_V < 4$ \\
    Slope Dev. ($\delta$) & Gaussian & $\mu=-0.35, \sigma=0.2$ \\
    UV Bump ($B$) & Gaussian & $\mu=2.27, \sigma=1.0$ \\
    Ionisation ($\log U$) & Uniform & $-3.5 < \log U < 0$ \\
    Escape Fraction ($f_{\text{esc}}$) & Log-Uniform & $0.001 < f_{\text{esc}} < 0.5$ \\
    Spec. Calibration & Gaussian & $\mu=1.0, \sigma=0.2$ \\ \hline
    \end{tabular}%
}
\caption{Adopted parameters and priors for the SED-fitting of NIRSpec spectroscopy and NIRCam photometry with Bagpipes.}
\label{tab:sed_priors}
\end{table}

\subsection{Derivation of Chemical Abundances}
\label{sec:chemical_abund}

The chemical abundances of carbon, nitrogen and oxygen are the central focus of this work, since they encode the enrichment history of our emission-line-selected population. The diagnostic lines involved do not arise in a single, uniform gas phase: JWST spectroscopy has shown that the ionised ISM of high-redshift galaxies is both denser than in local systems \citep{isobe2023ne} and stratified, with high-ionisation rest-UV lines tracing denser, more highly ionised gas than rest-optical diagnostics \citep{ji2024ne,topping25}. Robust abundance determinations therefore require each ion to be assigned the temperature and density of the zone in which it resides \citep{martinez2025}. We outline here our procedure for deriving accurate C, N, O, and Ne abundances from the measured emission lines, alongside corrections for dust attenuation and for unseen ionisation states. 

We note that because abundance measurements depend non-linearly on various line ratios, we derive associated uncertainties through bootstrap resampling. We draw 1000 realisations of the relevant line fluxes from their best-fit model posteriors, and for each run the full procedure outlined in the following sections. The final C, N, O, and Ne abundances and their uncertainties are derived from the median and semi-difference of the 16th and 84th percentiles of the resulting distribution.

\subsubsection{Dust Correction}
\label{sec:dustcorr}

To begin with, we verify and correct for the presence of any dust, as measured from multiple Balmer line ratios and comparing to Case B recombination ratios \citep{osterbrock06}. We use all available Balmer lines with SNR $\geq 5$, namely H$\gamma$, H$\delta$, H9, H10, and H$\beta$, and normalise according to the latter. H$\epsilon$ and H8 are excluded as they are blended, while H$\alpha$ is not considered due to it being covered only out to $z \simeq 7.5$ ($\sim35\%$ of our sample lies at $z > 7.5$) and our desire to maintain a consistent set of diagnostics. Each line yields an independent $A_V$ estimate assuming the \citet{cardelli1989} law with $R_V = 3.1$; the final value is the inverse-variance weighted mean, and fluxes are corrected as $F_\mathrm{int}(\lambda) = F_\mathrm{obs}(\lambda)\,10^{\,0.4\,k(\lambda)\,A_V/R_V}$. Considering the faintness of these lines, which are beyond the typical depths of individual spectra, we perform the dust corrections on the high SNR composite spectra rather than the individual spectra going into a given stack. We note that this $A_V$ traces the nebular extinction towards the ionised gas and is distinct from the stellar continuum extinction inferred by SED-fitting with Bagpipes.

\subsubsection{Multi-Zone Electron Temperature and Density}
\label{sec:tene}


Neglecting the multi-ionisation phase of ISM gas can significantly bias abundance measurements, in particular at high-redshift where collisional de-excitation becomes an important consideration in the densest regions. We therefore require a temperature and density for each line-emitting zone, and adopt the multi-zone scheme of \citet{martinez2025}.

The electron temperature of an ionisation zone is set by the ratio of a temperature-sensitive auroral line to a brighter nebular line of the same ion. These auroral lines are intrinsically faint, but by stacking $R\sim1000$ spectra we recover the [O\,\textsc{iii}]\,$\lambda4363$ \AA\ line in all of our composite spectra at $>8\sigma$ (see Sections~\ref{sec:full_stack} and \ref{sec:binned}), yielding a direct electron temperature measurement for each composite. Using the framework outlined in \citet{martinez2025}, we describe the ionised gas by three electron temperatures ($T_e^{\rm high}$, $T_e^{\rm int}$ and $T_e^{\rm low}$) tracing high, intermediate, and low electron density zones. The high-ionisation temperature is measured directly from the [O\,\textsc{iii}]\,$\lambda 4363$ / [O\,\textsc{iii}]$\lambda\lambda 4960,5008$ ratio with \textsc{PyNeb} \citep{pyneb}, adopting the emissivities of \citet{storeyzeippen2000}. The intermediate- and low-ionisation temperatures, which our data do not constrain directly, are inferred from $T_e^{\rm high}$ through the relations of \citet{garnett1992}:

\begin{align}
  T_e^{\rm int} &= 0.83\,T_e^{\rm high} + 1700 \mathrm{K}, \label{eq:te_int}\\
  T_e^{\rm low} &= 0.70\,T_e^{\rm high} + 3000 \mathrm{K}. \label{eq:te_low}
\end{align}

Each ion is then assigned the temperature of the zone from which its emission predominantly arises.
Furthermore, following \citet{martinez2025} we also characterise the density of each ionisation zone through $n_e$-sensitive diagnostics, and apply the ensuing constraints to the emission lines emanating from those zones. The low-ionisation density ($n_{e,\mathrm{low}}$) is derived from [S\,\textsc{ii}]\,$\lambda 6718 / \lambda 6732$ where covered, or otherwise from the rest-UV [Si\,\textsc{iii}]\,$\lambda 1883 / \lambda 1892$ doublet; the intermediate density ($n_{e,\mathrm{mid}}$) from C\,\textsc{iii}]\,$\lambda 1907 / \lambda 1909$; and the high-ionisation density ($n_{e,\mathrm{high}}$) from N\,\textsc{iv}]\,$\lambda 1483 / \lambda 1486$, all computed with \textsc{PyNeb}.
Where a density diagnostic is unavailable or falls below our SNR threshold, we instead utilise redshift-dependent trends of electron densities, evaluated at the median redshift of each composite:

\begin{align}
  n_{e,\mathrm{low}}  &= 54\,(1+z)^{1.2\pm0.4}                  && \text{\citep{abdurrouf2024}},   \label{eq:ne_low}\\
  n_{e,\mathrm{mid}}  &= 1.11\times10^{3}\,(1+z)^{1.93\pm0.08}  && \text{\citep{martinez2025}},    \label{eq:ne_mid}\\
  n_{e,\mathrm{high}} &= 5.40\times10^{3}\,(1+z)^{1.62\pm0.12}  && \text{\citep{martinez2025}},    \label{eq:ne_high}
\end{align}

The temperature and density prescriptions for each ion are summarised in Table~\ref{tab:abund_zones}.

\subsubsection{Direct Oxygen Abundance}
\label{sec:direct_metallicity}

We now describe our methodology for deriving a direct $T_e$-based oxygen abundance for each composite, using the electron temperature and density framework outlined in Section~\ref{sec:tene}. In the H\,\textsc{ii} regions of star-forming galaxies oxygen is present almost entirely as O$^+$ and O$^{2+}$; as such we evaluate O$^{2+}$/H$^+$ at $T_e^{\rm high}$ and O$^+$/H$^+$ at $T_e^{\rm low}$, although note that the hard radiation fields of high-redshift systems keep the bulk of the gas-phase oxygen in the O$^{2+}$ state.

With the electron temperatures and densities fixed, and the dust-corrected fluxes normalised to H$\beta$, we measure ionic abundances with \textsc{PyNeb}. O$^{2+}$/H$^+$ is derived from [O\,\textsc{iii}]\,$\lambda 5008$ and O$^+$/H$^+$ from [O\,\textsc{ii}]\,$\lambda\lambda 3726,3729$, and the total oxygen abundance is the sum of the two ionic states,
\begin{equation}
  12 + \log(\mathrm{O/H}) = 12 + \log_{10}\left(\frac{\mathrm{O^+}}{\mathrm{H^+}} + \frac{\mathrm{O^{2+}}}{\mathrm{H^+}}\right),
\end{equation}
with higher ionisation states (O$^{3+}$ and above) neglected. We note that for all of the composite spectra constructed in Sections~\ref{sec:full_stack} and \ref{sec:binned}, the O$^{2+}$ ion dominates the total oxygen abundance with $\mathrm{O^{2+}}/\mathrm{O_{tot}}$ ranging between
$\sim$85-95 per cent. For comparison, we also derive strong-line oxygen abundances using the calibrations of \citet{sanders25}, solving simultaneously for $12+\log(\mathrm{O/H})$ from the O3\,$\equiv[\mathrm{O\,\textsc{iii}}]\,\lambda5008/\mathrm{H}\beta$, O2\,$\equiv[\mathrm{O\,\textsc{ii}}]\,\lambda\lambda3726,3729/\mathrm{H}\beta$, R23\,$\equiv([\mathrm{O\,\textsc{iii}}]\,\lambda\lambda4960,5008+[\mathrm{O\,\textsc{ii}}])/\mathrm{H}\beta$ and O32\,$\equiv[\mathrm{O\,\textsc{iii}}]\,\lambda5008/[\mathrm{O\,\textsc{ii}}]\,\lambda\lambda3726,3729$ ratios.
The two metallicity estimates agree across all composites to within the intrinsic scatter of the strong-line calibrations (see Sections~\ref{sec:full_stack} and \ref{sec:binned}), although the strong-line values lie systematically lower, with a median offset of $0.12$~dex and a maximum of $0.31$~dex ($\simeq 2\sigma$).

\subsubsection{Nitrogen-to-Oxygen Ratio}
\label{sec:NO}

Accurate nitrogen-to-oxygen (N/O) derivations require accounting for nitrogen and oxygen gas residing across multiple ionisation states, with the fraction in each state set by the ionisation parameter of the gas. We infer the ionisation parameter from the observed $\mathrm{O32}$ 
ratio rather than from e.g., higher-ionisation C\,\textsc{iv}/C\,\textsc{iii}] ratios, which have been shown in high-redshift sources as complex P-Cygni profiles that incorporate contributions from both ISM gas, stellar winds, and outflowing gas (e.g., \citealt{glazer2025,marqueschaves2026}).
At the ionisation parameters typical of our sample ($\log U \sim -2.75$ to $-3.3$), a significant fraction of nitrogen resides in $\mathrm{N^{2+}}$ and $\mathrm{N^{3+}}$, measurable through the UV lines N\,\textsc{iii}]\,$\lambda\lambda1749,1752$ and N\,\textsc{iv}]\,$\lambda\lambda1483,1486$, while the oxygen is dominated by $\mathrm{O^{2+}}$ (traced by $[\mathrm{O\,\textsc{iii}}]\,\lambda\lambda4960,5008$). The intermediate-ionisation $\mathrm{N^{2+}}$ is evaluated at $T_e^{\rm int}$ and $n_{e,\mathrm{mid}}$, while the high-ionisation $\mathrm{N^{3+}}$ is evaluated at $T_e^{\rm high}$ and $n_{e,\mathrm{high}}$; Table~\ref{tab:abund_zones}.

Importantly, $(\mathrm{N^{2+}} + \mathrm{N^{3+}})/\mathrm{O^{2+}}$ does not by itself equal to total N/O, as it neglects contributions from other ionisation states. An ionisation correction factor (ICF) is required to recover the total elemental abundance ratio, therefore we adopt the ICFs of \citet{martinez2025} which depend on the ionisation parameter, metallicity and electron density of the gas, each of which are inferred directly from the composite spectra as described in previous sections.
We do not use prescriptions requiring $[\mathrm{N\,\textsc{ii}}]\,\lambda6585$, since this line falls beyond the NIRSpec detector at $z \gtrsim 7.5$ and is covered in only approximately 65 per cent of our fiducial sample.

\subsubsection{Carbon-to-Oxygen Ratio}
\label{sec:CO}

The carbon-to-oxygen (C/O) ratio serves as an additional constraint for the underlying enrichment mechanisms and provides a baseline for interpreting the origin of elevated N/O. Both C and O are produced predominantly by massive stars on short timescales, so C/O is a relatively stable tracer of short-lived massive star yields, while longer timescales are typically required for substantial nitrogen enrichment.
As such, measured together, C/O and N/O therefore serve as diagnostics of the recent star formation and the underlying stellar populations \citep[e.g.,][]{henry2000, berg2019, perezmontero2021}. Carbon abundance is derived from C\,\textsc{iii}]\,$\lambda\lambda1907,1909$ using $T_e^{\rm int}$ and $n_{e,\mathrm{mid}}$. Although C\,\textsc{iv}\,$\lambda\lambda1548,1550$ would in principle constrain the higher-ionisation C$^{3+}$, this doublet is affected by strong absorption in high-redshift galaxies. We therefore only use C\,\textsc{iii}] and adopt the C$^{2+}$/O$^{2+}$ ICF of \citet{martinez2025} that accounts for carbon in higher ionisation states.




\subsubsection{Neon-to-Oxygen Ratio}
\label{sec:NeO}

Neon and oxygen are both $\alpha$-elements synthesised by the same massive stars and ejected by CCSNe in near constant proportion, with no significant secondary or delayed contribution. As a result, the neon-to-oxygen (Ne/O) ratio serves as an $\alpha$-element baseline against which N/O can be compared. $\mathrm{Ne/O}$ is derived entirely from high-ionisation tracers: Ne$^{2+}$/H$^+$ from [Ne\,\textsc{iii}]\,$\lambda 3869$ and O$^{2+}$/H$^+$ from [O\,\textsc{iii}]\,$\lambda\lambda 4960,5008$, both computed in the high-ionisation zone at $T_e^{\rm high}$ and $n_{e,\mathrm{mid}}$.
Because Ne$^+$ is not accessible in our NIRSpec spectra, we apply the ICF of \citet{izotov2006}.

The redshift fitting of Section~\ref{sec:sample}, the line fitting of Section~\ref{sec:linefitting} and the abundance derivations described in this section are implemented in \textsc{jwspecfit}\footnote{\url{https://github.com/raunaq-rai/jwspecfit}\label{fn:jwspecfit}}, a publicly available Python package; full implementation details are documented in the repository.

\begin{table}
\centering
\begin{tabular}{llll}
    \hline
    Ion & Line(s) & $T_e$ & $n_e$ \\
    \hline
    $\mathrm{O^{+}/H^{+}}$   & [O\,\textsc{ii}]\,$\lambda\lambda3726,3729$  & low  & low ([S\,\textsc{ii}]) \\
    $\mathrm{O^{2+}/H^{+}}$  & [O\,\textsc{iii}]\,$\lambda5008$             & high & mid (C\,\textsc{iii}]) \\
    $\mathrm{Ne^{2+}/H^{+}}$ & [Ne\,\textsc{iii}]\,$\lambda3869$            & high & mid (C\,\textsc{iii}]) \\
    $\mathrm{C^{2+}/H^{+}}$  & C\,\textsc{iii}]\,$\lambda\lambda1907,1909$  & int  & mid (C\,\textsc{iii}]) \\
    $\mathrm{C^{3+}/H^{+}}$  & C\,\textsc{iv}\,$\lambda\lambda1548,1550$     & high & high (N\,\textsc{iv}]) \\
    $\mathrm{N^{2+}/H^{+}}$  & N\,\textsc{iii}]\,$\lambda\lambda1749,1752$  & int  & mid (C\,\textsc{iii}]) \\
    $\mathrm{N^{3+}/H^{+}}$  & N\,\textsc{iv}]\,$\lambda\lambda1483,1486$   & high & high (N\,\textsc{iv}]) \\
    \hline
\end{tabular}
\caption{Electron temperature and density zones adopted for each ionic abundance, following the three-zone framework of \citet{martinez2025}.
$T_e(\mathrm{high}) \equiv T_e(\mathrm{O^{2+}})$ is measured from [O\,\textsc{iii}]\,$\lambda4363/[\mathrm{O\,\textsc{iii}}]\,\lambda\lambda4960,5008$; the intermediate ($T_e^{\rm
int}$) and low ($T_e^{\rm low}$) temperatures follow from the \citet{garnett1992} relations (Equations~\ref{eq:te_int} and \ref{eq:te_low}). The
high-ionisation ions O$^{2+}$ and Ne$^{2+}$ are evaluated at $T_e^{\rm high}$, and the intermediate-ionisation ions C$^{2+}$ and N$^{2+}$ at $T_e^{\rm int}$.
The low density is measured from [S\,\textsc{ii}]\,$\lambda6718/\lambda6732$ (or rest-UV [Si\,\textsc{iii}]\,$\lambda1883/\lambda1892$), the intermediate
density from C\,\textsc{iii}]\,$\lambda1907/\lambda1909$, and the high density from N\,\textsc{iv}]\,$\lambda1483/\lambda1486$; where a diagnostic is unavailable, we utilise redshift-dependent relations (Equations~\ref{eq:ne_low}--\ref{eq:ne_high}). Note, for O$^{2+}$ ($35.1$--$54.9$~eV) and Ne$^{2+}$ ($41.0$--$63.5$~eV) we adopt $n_{e,\mathrm{mid}}$ from C\,\textsc{iii}] ($24.4$--$47.9$~eV).
}
\label{tab:abund_zones}
\end{table}

\section{The Average Spectrum of \texorpdfstring{$\bm{6 \leq \lowercase{z} \leq 10}$}{6 ≤ z ≤ 10} Galaxies}
\label{sec:full_stack}

\begin{figure*}
  \centering
  \includegraphics[width=\textwidth]{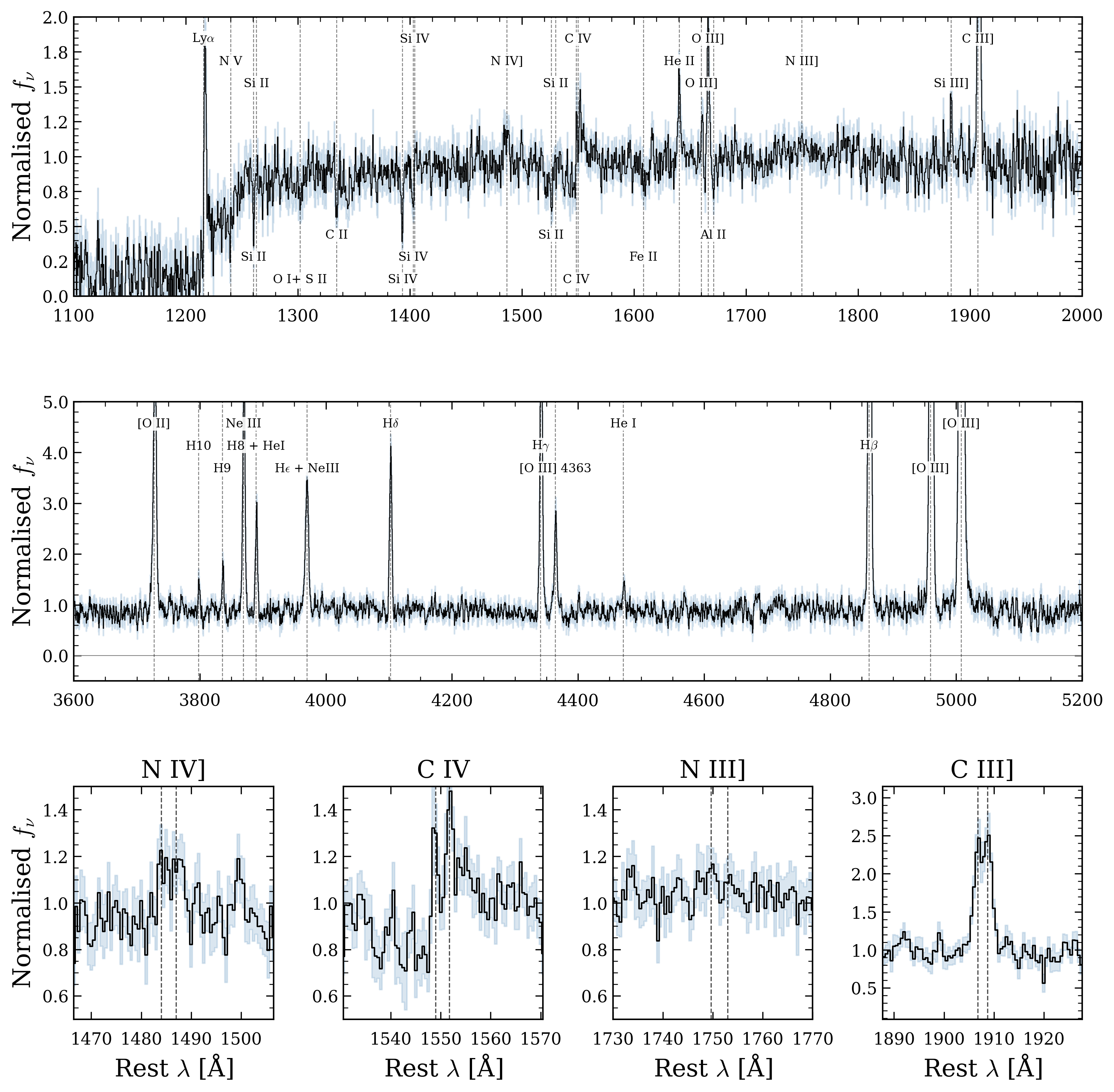}
    \caption{Composite rest-frame spectrum of the 135 galaxies satisfying our selection criteria ($6 \leq z \leq 10$, $-21 < M_{\rm UV} <-19$, $\mathrm{SNR}_{\mathrm{UV}} > 0.2$). \textit{Top:} Rest-frame UV ($1100$--$1800$\AA), revealing interstellar absorption features and faint nebular emission lines that are typically undetected in the individual spectra of galaxies at this luminosity and redshift. \textit{Middle:} Rest-frame optical ($3600$--$5200$\AA) covering [O\,\textsc{ii}], the higher-order Balmer series, [O\,\textsc{iii}]\,$\lambda4363$, and [O\,\textsc{iii}]\,$\lambda\lambda4960,5008$. \textit{Bottom:} Zoom-in panels on the N\,\textsc{iv}]\,$\lambda\lambda1483,1486$, C\,\textsc{iv}\,$\lambda\lambda1548,1550$, N\,\textsc{iii}]\,$\lambda\lambda1749,1752$ and C\,\textsc{iii}]\,$\lambda\lambda1907,1909$ doublets. This composite was constructed by taking the median of the normalised spectra, and is therefore representative of the typical spectral properties of the population.}
  \label{fig:full_stack}
\end{figure*}

\begin{table*}
\centering
\renewcommand{\arraystretch}{1.3}
\begin{tabular}{@{}lr@{\hspace{3.5em}}lr@{\hspace{3.5em}}lr@{}}
\toprule
Line & $W_0$ (\AA) & Line & $W_0$ (\AA) & Line & $W_0$ (\AA) \\
\midrule
Si\,{\sc ii}~$\lambda$1260$^{*}$        & $0.5^{+0.2}_{-0.2}$   & O\,{\sc iii}]~$\lambda$1666            & $2.3^{+0.2}_{-0.2}$   & H$\gamma$                        & $34.2^{+0.8}_{-0.7}$ \\
O\,{\sc i}$+$Si\,{\sc ii}~$\lambda$1303$^{*}$ & $0.6^{+0.2}_{-0.3}$ & Al\,{\sc ii}~$\lambda$1672$^{*}$   & $0.6^{+0.2}_{-0.2}$   & {[O\,{\sc iii}]~$\lambda$4363}   & $11.5^{+0.4}_{-0.4}$ \\
C\,{\sc ii}~$\lambda$1334$^{*}$         & $0.6^{+0.2}_{-0.2}$   & N\,{\sc iii}]~$\lambda$1749            & $0.3^{+0.1}_{-0.1}$   & He\,{\sc i}~$\lambda$4472        & $2.8^{+0.3}_{-0.3}$ \\
Si\,{\sc iv}~$\lambda$1393$^{*}$        & $1.1^{+0.2}_{-0.2}$   & N\,{\sc iii}]~$\lambda$1752            & $0.1^{+0.1}_{-0.1}$   & He\,{\sc ii}~$\lambda$4687       & $1.6^{+0.3}_{-0.3}$ \\
Si\,{\sc iv}~$\lambda$1403$^{*}$        & $0.6^{+0.2}_{-0.2}$   & C\,{\sc iii}]~$\lambda$1907            & $3.0^{+0.3}_{-0.3}$   & H$\beta$                         & $86.5^{+1.7}_{-1.8}$ \\
N\,{\sc iv}]~$\lambda$1483              & $0.5^{+0.1}_{-0.1}$   & C\,{\sc iii}]~$\lambda$1909            & $3.3^{+0.3}_{-0.3}$   & {[O\,{\sc iii}]~$\lambda$4960}   & $173.2^{+3.1}_{-3.2}$ \\
N\,{\sc iv}]~$\lambda$1486              & $0.6^{+0.1}_{-0.1}$   & {[O\,{\sc ii}]~$\lambda\lambda$3726,3729} & $34.5^{+1.0}_{-1.0}$ & {[O\,{\sc iii}]~$\lambda$5008} & $527.7^{+9.1}_{-9.0}$ \\
Si\,{\sc ii}~$\lambda$1526$^{*}$        & $1.1^{+0.3}_{-0.3}$   & H10                                    & $2.0^{+0.3}_{-0.3}$   & He\,{\sc i}~$\lambda$5877        & $12.4^{+0.6}_{-0.6}$ \\
C\,{\sc iv}~$\lambda$1548               & $0.4^{+0.2}_{-0.2}$   & H9                                     & $3.5^{+0.3}_{-0.3}$   & {[O\,{\sc i}]~$\lambda$6302}     & $4.1^{+0.7}_{-0.7}$ \\
C\,{\sc iv}~$\lambda$1550               & $1.0^{+0.2}_{-0.2}$   & {[Ne\,{\sc iii}]~$\lambda$3869}        & $25.0^{+0.6}_{-0.6}$  & H$\alpha$                        & $358.7^{+6.2}_{-6.2}$ \\
Fe\,{\sc ii}~$\lambda$1608$^{*}$        & $0.9^{+0.2}_{-0.2}$   & H8$ + $He\,{\sc i}~$\lambda$3889         & $9.1^{+0.4}_{-0.4}$   & {[N\,{\sc ii}]~$\lambda$6585}    & $8.9^{+1.0}_{-1.0}$ \\
He\,{\sc ii}~$\lambda$1640              & $1.7^{+0.2}_{-0.2}$   & H$\epsilon + $[Ne\,{\sc iii}]~$\lambda$3968 & $15.9^{+0.4}_{-0.4}$ & {[S\,{\sc ii}]~$\lambda$6718} & $3.1^{+1.0}_{-1.0}$ \\
O\,{\sc iii}]~$\lambda$1661             & $0.6^{+0.1}_{-0.1}$   & H$\delta$                              & $15.3^{+0.4}_{-0.4}$  & {[S\,{\sc ii}]~$\lambda$6732}    & $3.2^{+0.7}_{-0.7}$ \\
\bottomrule
\end{tabular}
\caption{Rest-frame equivalent widths of the emission and absorption lines detected in the full-sample composite spectrum from Section~\ref{sec:full_stack}. Absorption features are marked with an asterisk ($^{*}$), and for these the absolute equivalent width is quoted.}
\label{tab:ew}
\end{table*}

With a statistical sample of galaxies with $R \sim 1000$ spectra in hand, we first examine the spectroscopic features that dominate the ISM and stellar signatures of $6 \leq z \leq 10$ galaxies. Figure~\ref{fig:full_stack} shows the resulting composite spectrum of the 135 galaxies in our fiducial sample, in which several features and key diagnostics central to this analysis are seen, and Table~\ref{tab:ew} lists the rest-frame equivalent widths of every emission and absorption line measured from it.

In addition to the strong [O\,\textsc{iii}]\,$\lambda\lambda4960,5008$ and H$\beta$ lines typically identified in individual spectra which allow for a robust redshift determination, the spectrum also reveals a suite of strong rest-optical lines that serve as powerful diagnostics for the ionisation state and metallicity of the ISM gas: in particular, [O\,\textsc{ii}]\,$\lambda\lambda3726,3729$ (blended), [Ne\,\textsc{iii}]\,$\lambda3869$, and [Ne\,\textsc{iii}]\,$\lambda3968$. Measured line ratios of log\,$\mathrm{O32}=0.93\pm0.02$ and log\,$\mathrm{R23}=0.95\pm0.01$ point to hard ionising spectra and low-metallicity conditions that contrast with the local Universe and are consistent with typical measurements of $z>5$ systems \citep{cameron2023b,robertsborsani2024}.

More critical to this work, the composite spectrum reveals prominent auroral lines from O\,\textsc{iii}]\,$\lambda\lambda1661,1666$ and [O\,\textsc{iii}]\,$\lambda4363$, which allow for a robust, electron temperature-based O abundance measurement as outlined in Section~\ref{sec:direct_metallicity}. The clear detection of these lines, given their relative faintness and close proximity to multiple lines, serves as a clear demonstration of the resolving power of especially deep $R\sim1000$ spectroscopy to pick up ISM diagnostics that are beyond the capabilities of prism spectroscopy. Similarly, the full Balmer series from H$\alpha$ through H$\beta$, H$\gamma$ and H$\delta$, to H10 emission is also detected, providing multiple independent constraints on interstellar dust reddening via the Balmer decrement, as well as diagnostics of recent star formation rate and the production rate of ionising continuum photons, $\xi_{\rm ion}$ (e.g., \citealt{glazer2025}).


Remarkably, the spectrum also reveals a clear detection of the nebular N\,\textsc{iv}]\,$\lambda\lambda1483,1486$ doublet, a high-ionisation feature requiring photons with energies $>47.4$~eV. The emergence of N\,\textsc{iv}] in a composite spectrum representative of the moderately luminous, emission-line galaxy at $z>6$ is perhaps unexpected, owing to its apparent rarity across the bulk population of both low- and high-redshift sources studied with JWST until now \citep{mingozzi22,schaerer2024}. The clear detection of the line shown here establishes it as a ubiquitous and generic feature of our sample, rather than a rare and exotic feature exclusive to a small number of extreme and luminous outliers. The N\,\textsc{iii}]\,$\lambda\lambda1749,1752$ line, whose emission requires ionising photons of $>29.6$~eV, on the other hand is only measured at $\sim2.9\sigma$, suggesting the majority has been ionised by a hard radiation field into N\,\textsc{iv}]. For comparison, [N\,\textsc{ii}]\,$\lambda6585$ is clearly detected, which likely points to a clear stratification of the ISM gas \citep{isobe2023ne,topping25}.


C\,\textsc{iii}]\,$\lambda\lambda1907,1909$ and C\,\textsc{iv}\,$\lambda\lambda1548,1550$ doublets are also clearly seen in the composite. Strong carbon emission has been detected in individual star-forming systems (both luminous and faint) and AGN at $z > 6$ \citep[e.g.,][]{bunker2023gnz11, topping2024, castellano2024,arevalogonzalez26}, as well as in stacked average spectra at $5<z<10$ \citep[e.g.,][]{robertsborsani2024,glazer2025,hu2024,hayes2024,umeda2026} and at $z>10$ \citep{robertsborsani2026}. Its clear detection in our $R\sim1000$ composite reaffirms that strong carbon emission is a generic property of $6 \leq z \leq 10$ star-forming populations. Producing C$^{3+}$ requires photons of $>47.9$~eV, therefore, as with N\,\textsc{iv}], C\,\textsc{iv} traces an especially hard ionising field. Locally, such nebular C\,\textsc{iv} emission is almost exclusive to extremely metal-poor, young, and highly star-forming dwarfs and Lyman-continuum leakers \citep{senchyna2017, schaerer2022, saxena2022}. At $z>6$, nebular C\,\textsc{iv} has become increasingly prevalent, although largely confined to individual UV-luminous or strongly star-forming systems \citep[e.g.,][]{stark2015, mainali2017, topping2021, castellano2024,robertsborsani2026}. The ubiquity of C\,\textsc{iv} in our composite once more indicates that the feature and the extreme ionising conditions it traces are typical of the $6 \leq z \leq 10$ population, rather than confined to extreme systems. The C\,\textsc{iv}\,$\lambda\lambda$1548,1550 profile additionally shows a clear P-Cygni component with extended (redshifted) emission and broader (blueshifted) absorption, characteristic of strong stellar winds \citep{castor1975}. The feature is attributed to young, massive OB-type stars \citep{shapley2003, chisholm2019, toppingCN}, so its presence suggests a spectrum dominated by relatively young stellar populations. The broad absorption profile likely also carries a combination of photospheric and interstellar contributions \citep{chisholm2019}.

He\,\textsc{ii}\,$\lambda1640$ emission is also revealed by the composite. This recombination line requires photon energies exceeding $54.4$\,eV, making it one of the clearest tracers for an especially hard ionising spectrum and frequently invoked as a signature of AGN activity. The detection of the line in this composite corroborates the hard ionising field inferred from the O32 ratio above, necessary to produce N\,\textsc{iv}] and C\,\textsc{iv}, while the exclusion of AGN from our sample in Section~\ref{sec:sample} suggests the line is powered by stellar origins and represents a generic feature of star-forming galaxies at low metallicities \citep[e.g.,][]{senchyna2017, berg2019, saxena2020}.


Turning to chemical abundances, our full-sample composite is characterised by a metallicity of $12 + \log(\mathrm{O/H}) = 7.79^{+0.03}_{-0.03}$, as measured from the [O\,\textsc{iii}]\,$\lambda4363$ line following the procedure outlined in Section~\ref{sec:chemical_abund}. This is characteristic of the low-metallicity gas enabling hard radiation fields in star-forming systems at $6 \leq z \leq 10$, where galaxies have had less than $\sim 1$ Gyr to enrich their ISM \citep{curti2024}. Notably, the full-sample composite is also described by a supersolar nitrogen-to-oxygen ratio of $\log(\mathrm{N/O}) = -0.57^{+0.09}_{-0.10}$, well above the subsolar values measured in local dwarf galaxies \citep{izotov2006, pilyugin2012, berg2019}. Remarkably, this enrichment is already in place at such low metallicity and so early in cosmic time, a result whose origin we examine in detail in Sections~\ref{sec:binned} and \ref{sec:discussion}. The fact such an elevated N/O ratio is recovered in a median composite, where a handful of individual outliers cannot drive this, suggests that nitrogen enhancement is a common property of the population our sample probes rather than a peculiarity of extreme objects. It follows that existing prism samples, assembled from the most luminous line emitters, understate the prevalence of nitrogen enhancement. We note that this N/O ratio is derived from the UV nitrogen lines,
while repeating the measurement with the optical [N\,\textsc{ii}]\,$\lambda6585$ line instead yields a lower $\log(\mathrm{N/O}) = -1.17^{+0.06}_{-0.06}$, an offset of $\approx 0.6$~dex between the UV- and optical-based values. A similar discrepancy is reported by \citet{umeda2026} in their stacked $R \sim 1000$ spectra at $z \simeq 4.5-10$, where N\,\textsc{iv}]-based N/O is supersolar ($\log(\mathrm{N/O})_{\rm UV} = -0.20 \pm 0.24$) while the [N\,\textsc{ii}]-based value sits on the local relation ($\log(\mathrm{N/O})_{\rm Opt} = -1.59^{+0.11}_{-0.10}$); they suggest this may arise from a localised, highly ionised nitrogen-enriched component to which the UV lines are preferentially sensitive.
In contrast, the C/O ratio of our full-sample composite is subsolar with $\log(\mathrm{C/O}) = -0.80^{+0.03}_{-0.03}$ and lies only moderately below ($\approx 0.08$~dex) the local metal-poor galaxy sequence at comparable O/H \citep[although consistent within uncertainties;][]{nicholls2017, berg2016, berg2019}.
The combination of supersolar N/O and subsolar C/O is consistent with values associated with a number of luminous, nitrogen-enhanced sources at high-redshift \citep{marques2024,schaerer2024} and points to a ubiquity amongst the $z>6$ line-emitting population. We also measure $\log(\mathrm{Ne/O}) = -0.78^{+0.01}_{-0.01}$, which is consistent with the solar value and with local star-forming galaxies, suggesting the composite and its abundances are not subject to $\alpha$-enhancement.



In the rest-frame UV, the composite also reveals prominent absorption from low-ionisation features, such as Si\,\textsc{ii}\,$\lambda1260$, O\,\textsc{i}$+$Si\,\textsc{ii}\,$\lambda1303$, C\,\textsc{ii}\,$\lambda1334$, Si\,\textsc{ii}\,$\lambda1526$, Fe\,\textsc{ii}\,$\lambda1608$, and Al\,\textsc{ii}\,$\lambda1672$, as well as from high-ionisation Si\,\textsc{iv}\,$\lambda\lambda1393,1403$, which together probe the kinematics and covering fraction of the neutral and ionised gas \citep{shapley2003}.
Following Equations~1 and~2 of \citet{vasan2026}, we model the absorption features in both ionisation zones with covering fractions and velocity offsets relative to the systemic velocity set by the nebular emission gas. Adopting Gaussian fits for the low-ionisation lines yields a central covering fraction $C_f = 0.34 \pm 0.04$ at a velocity consistent with systemic, $v_\mathrm{cen} = -1 \pm 22$\,km\,s$^{-1}$, with individual transitions spanning $C_f = 0.26$--$0.37$ and $v_\mathrm{cen} = -90$\,km\,s$^{-1}$ to $+43$\,km\,s$^{-1}$, indicative of both blueshifted and redshifted gas relative to systemic.
The high-ionisation Si\,\textsc{iv} gas, given its nature as a resonant doublet, is instead characterised by a single covering fraction and velocity offset which come to $C_f = 0.47 \pm 0.05$ and $v_\mathrm{cen} = -155 \pm 33$\,km\,s$^{-1}$, respectively.
Both covering fractions should be regarded as sample-averaged lower limits, since stacking galaxies with a range of outflow velocities broadens and shallows the trough, and since $C_f$ is recovered under the assumption that the lines are optically thick.
The low-ionisation range lies at the lower end of values measured for six individual $z = 5-9$ galaxies by \citet{vasan2026} ($C_f = 0.23-0.91$, mean $0.66$) and below those of $z = 7.2-10.6$ galaxies measured by \citet{nakane2026} ($C_f = 0.53$--$0.69$), while the Si\,\textsc{iv} value also lies below the high-ionisation covering fractions of the latter ($C_f = 0.61$--$0.92$). The combination of higher covering fractions and larger (plus systematically) blueshifted velocity offsets in the high-ionisation gas, compared to the values probed by the low-ionisation gas is notable; the findings are similar to those of \citet{vasan2026} and consistent with a scenario in which the high-ionisation gas preferentially traces star-formation-driven outflowing material that is more efficiently entrained than the low-ionisation gas, which lies closer to systemic velocities and, covering less of the continuum than the high-ionisation phase, reveals a more patchy neutral ISM.
Moreover, in a similar stacking analysis, \citet{glazer2025} compare blueshifted absorption in low-ionisation gas at $z \sim 7$ (with a mean of $-23 \pm 51$\,km\,s$^{-1}$) to blueshifts in the same gas seen at $z \simeq 3$--$5$, finding weaker outflow signatures in the former and attributing this to three possible causes: a lower bulk gas velocity, a more collimated outflow, or an outflow that is more highly ionised and therefore weaker in the low-ionisation transitions. Our measurements are most consistent with the last scenario, and point to a high prevalence of multi-phase outflows in our $6 \leq z \leq 10$ sample.

Lastly, we also identify a deep and broad absorption trough blueward of N\,\textsc{v}$\lambda$1240, which could (together with C\,\textsc{iv}) represent part of a P-Cygni wind signature from young and massive -- or even very massive -- stars \citep{marqueschaves2026,marqueschaves26a}. We note neither of the two VMS objects studied by \citet{marqueschaves2026} are present in this composite, owing to their extreme luminosities which exclude them from our selected sample. To verify this wind interpretation, we model the UV portion of the spectrum (with a focus on the aforementioned features) with both the Bagpipes code (as described in Section~\ref{sec:sed_fitting}) and the FICUS code \citep{saldanalopez23}. In the case of the latter, we assume the BPASS stellar templates (as for Bagpipes) and those from Starburst99 \citep{leitherer1999}, along with a 100\,$M_{\odot}$ IMF upper-mass cutoff and a constant star formation history, which is likely better suited to a composite spectrum whose effective star formation history is smooth. We find overall the spectral fits from both of these codes and each of these SSP templates are able to reproduce the UV spectrum (according to both $\chi^{2}$ metrics and visual inspection), and in particular fit both the C\,\textsc{iv} P-Cygni profile and the N\,\textsc{v} blueshifted trough reasonably well, indicating the wind features observed in our composites can be accounted for by standard massive-star populations under a canonical IMF.

\section{Linking Elemental Abundance Patterns to Recent Star Formation}
\label{sec:binned}

Having established from the full-sample composite that supersolar N/O and subsolar C/O are generic features of the underlying $6 \leq z \leq 10$ population, we now investigate whether this enrichment depends on the recent star formation history of that population. This consideration is especially pertinent at the high redshifts probed by our sample, where galaxies are increasingly found to undergo episodic bursts of stochastic star formation that can deviate markedly from their longer-term averages \citep[e.g.,][]{looser2025, endsley2025}.
To test this, we divide our galaxy sample by a measure of recent star formation (SF) and construct composite spectra with three aims: first, to establish whether measured equivalent widths of high-ionisation N\,\textsc{iv}], N\,\textsc{iii}], C\,\textsc{iv} and C\,\textsc{iii}] are sensitive to bursts of SF; second, to investigate how C, N, and O abundance ratios might vary as a function of recent burst timescale;
and third, where SNR permits, to search for spectral signatures of exotic and prompt enrichment channels, such as stellar winds from Wolf Rayet, very massive, or supermassive stars. Together, these connect the ubiquitous nitrogen and carbon lines seen in the full-sample composite, as well as their supersolar N/O and subsolar C/O abundance ratios, to the interplay of standard chemical enrichment frameworks and bursty star formation that dominate the ``typical'' star-forming source rather than requiring exotic enrichment channels. Although Section~\ref{sec:full_stack} showed that N/O estimates can vary according to the choice of tracer used, here we adopt UV-based N and C tracers, given the uniform coverage and \textit{relative} nature of our comparison.


\subsection{Star Formation Rate Excess}
\label{sec:birthrate}

\begin{figure}
\includegraphics[width=\columnwidth]{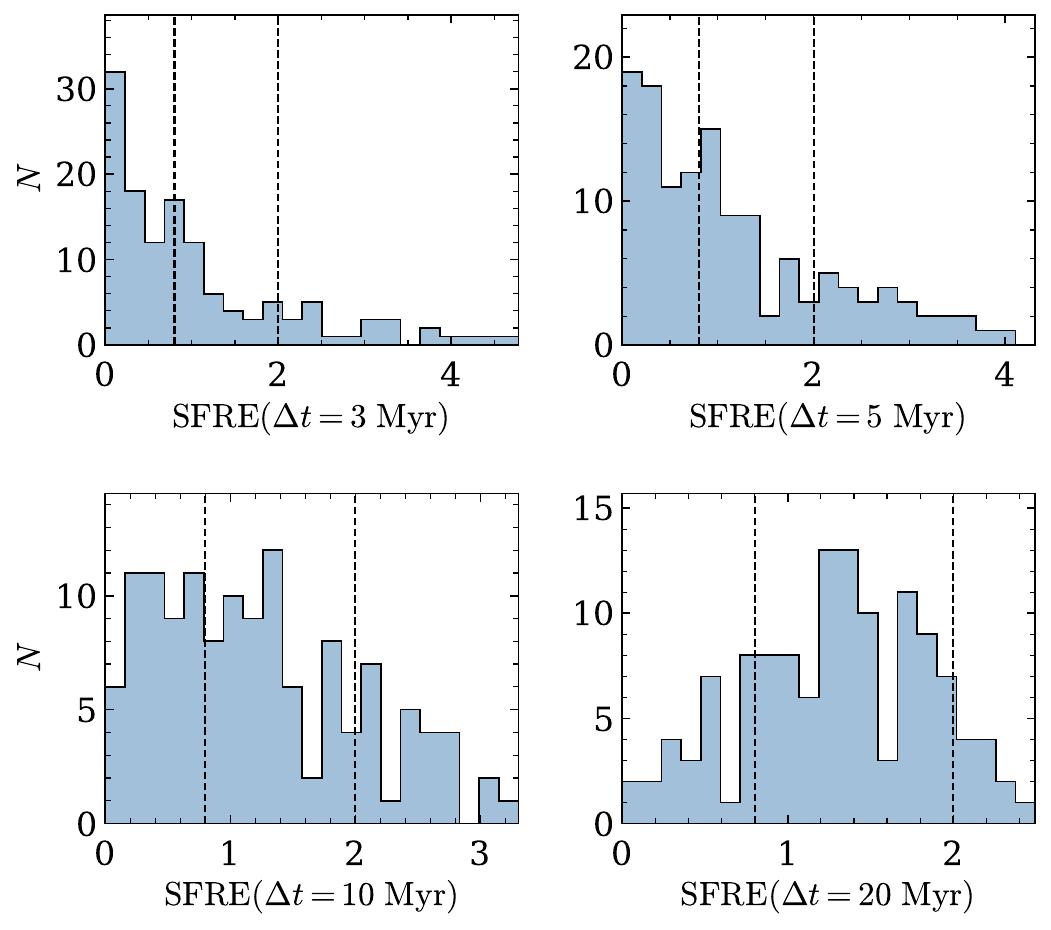}
\caption{Distribution of the star formation rate excess, SFRE, for each burst window ($\Delta t$) considered. Galaxies with SFRE$>$1 have formed a disproportionate fraction of their stellar mass within the last $\Delta t$, while SFRE$<$1 indicates a recent lull in star formation over the last $\Delta t$. Dashed black vertical lines mark the bin edges used to define ``lulling'', ``moderate'', and ``bursty'' composites.}
\label{fig:birthrate}
\end{figure}

We utilise the SFHs returned by the Bagpipes SED fits described in Section~\ref{sec:sed_fitting}, which provide an empirical handle on the recent SFR. From each SFH we define the star formation rate excess (SFRE), the ratio of a galaxy's recent SFR (defined over a timescale of choice, $\Delta t$) to its lifetime-averaged SFR, which quantifies whether it has experienced a recent upturn or lull in star formation:
  
\begin{equation}
    \mathrm{SFRE}(\Delta t) \;=\; \frac{\langle \mathrm{SFR}\rangle_{<\Delta t}}{\langle \mathrm{SFR}\rangle_{\mathrm{lifetime}}} \;=\; \frac{M_\star(<\Delta t)\,/\,\Delta t}{M_{\star,\mathrm{total}}\,/\,t_{\mathrm{gal}}},
    \label{eq:birthrate}
\end{equation}

where the numerator $\langle \mathrm{SFR}\rangle_{<\Delta t}$ is the mean star formation rate over the most recent $\Delta t$ and the denominator $\langle \mathrm{SFR}\rangle_{\mathrm{lifetime}}$ the mean over the galaxy's lifetime. $M_{\star,\mathrm{total}} = \int_0^{t_{\mathrm{gal}}} \mathrm{SFR}(t)\,\mathrm{d}t$ is the total stellar mass formed over the galaxy's lifetime, and $M_{\star}(< \Delta t) = \int_0^{\Delta t} \mathrm{SFR}(t)\,\mathrm{d}t$ is the mass formed within the most recent $\Delta t$. $t_{\mathrm{gal}}$ is the mass-weighted galaxy age returned by the SED fit.
A galaxy that has formed stars over the last $\Delta t$ at its lifetime-averaged rate has SFRE$=$1, while SFRE$>$1 indicates a recent upturn relative to that average, and SFRE$<$1 a recent decline, or ``lull''. We evaluate SFRE over timescales of $\Delta t = 3, 5, 10$ and $20$\,Myr and show the resulting SFRE$(\Delta t)$ distributions for each in Figure~\ref{fig:birthrate}. We construct three composite spectra per timescale, binning according to the following thresholds: lulling galaxies with SFRE$<$0.8, moderately star-forming galaxies with $0.8 \leq \mathrm{SFRE} < 2$, and extremely bursty galaxies with $\mathrm{SFRE} \geq 2$.
While star formation histories recovered on timescales as short as a few Myr carry an inherent degree of uncertainty, our analysis rests on the relative comparison of spectral properties across the SFRE distribution of a given burst window, and is therefore less sensitive to uncertainties across timescales.

In order to verify that SFRE$(\Delta t)$ is a meaningful tracer of bursty star formation, we compare it against spectroscopic indicators measured directly from the twelve composite spectra (three SFRE bins across four $\Delta t$ distributions), namely H$\beta$ equivalent width.
At $\Delta t = 3$~Myr we find H$\beta$ EW increases monotonically across the three SFRE bins, from $W_{\mathrm{H}\beta} = 63 \pm 3$\,\AA\ in the lulling composite, through $130 \pm 5$\,\AA\ in the moderate composite, to $179 \pm 8$\,\AA\ in the bursty composite, a factor of $\simeq 2.8\times$ increase from lulling to bursty. The same ordering holds at all other burst timescale windows, with $W_{\mathrm{H}\beta} \simeq 63$--$70$\,\AA\ measured for the lulling composites, $\simeq 76$--$130$\,\AA\ for the moderate composites, and $\simeq 119$--$179$\,\AA\ for the bursty composites, demonstrating a clear evolution in H$\beta$ line strength with SFRE that is not exclusive to one particular bursty timescale and which validates SFRE as an appropriate metric for our purposes.

\subsection{Characterising UV Emission Line Strengths}

\begin{figure}
    \centering
    \includegraphics[width=1\linewidth]{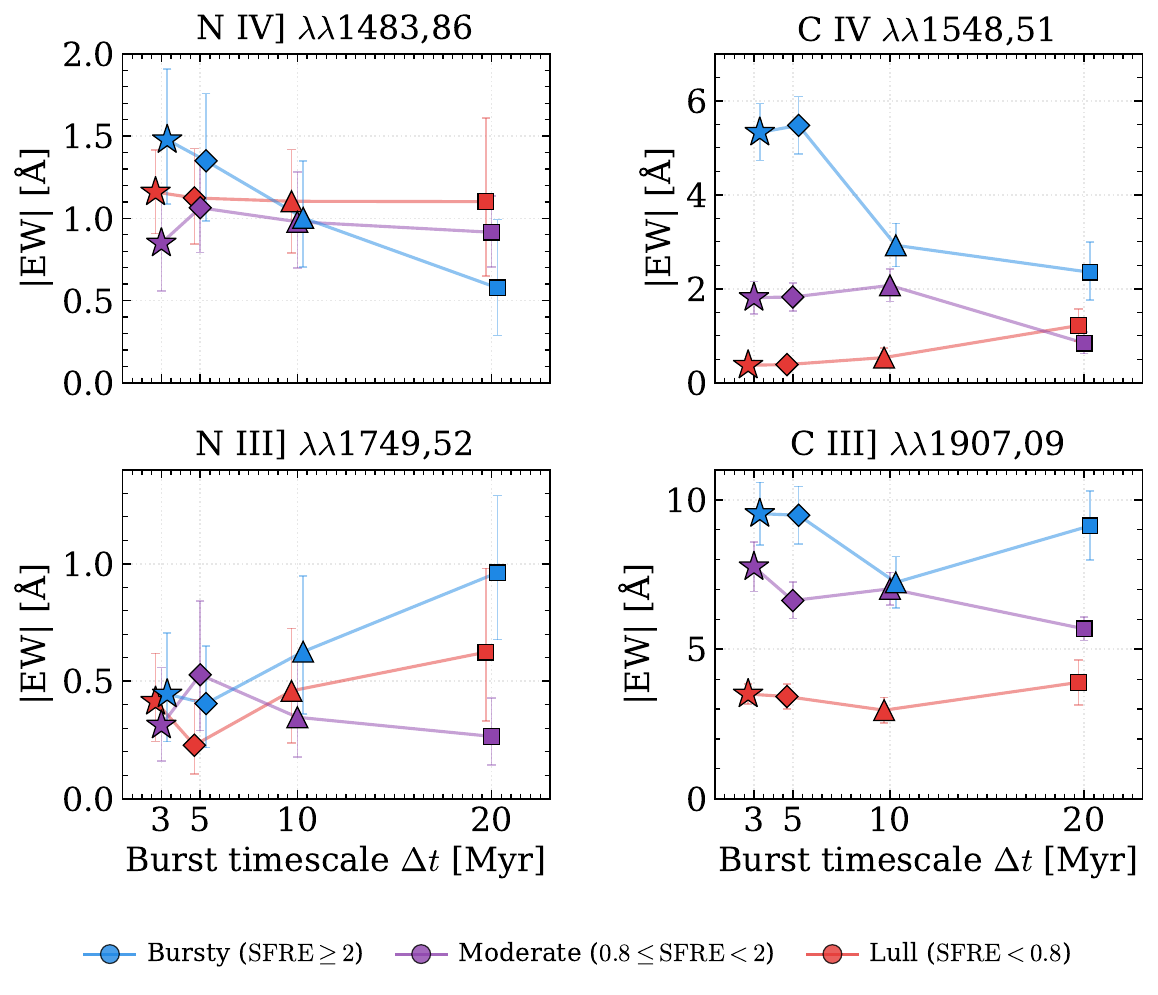}
    \caption{Rest-frame equivalent widths of N\,\textsc{iv}] (top left), N\,\textsc{iii}] (bottom left), C\,\textsc{iv} (top right), and C\,\textsc{iii}] (bottom right) as a function of burst timescale $\Delta t$ and SFRE. Burst timescales are denoted by different marker shapes: stars ($\Delta t = 3$~Myr), diamonds ($\Delta t = 5$~Myr), triangles ($\Delta t = 10$~Myr) and squares ($\Delta t = 20$~Myr). SFRE binning is indicated by different colours: lulling galaxies in red, moderate galaxies in purple, and bursty galaxies in blue. Error bars denote 1$\sigma$ uncertainties.}
    \label{fig:ew_vs_timescale}
\end{figure}

For each of our SFRE($\Delta t$) composites we measure the equivalent widths of C\textsc{iii}], C\,\textsc{iv}, N\,\textsc{iii}], and N\,\textsc{iv}], according to the methodology outlined in Section~\ref{sec:linefitting}. The resulting measurements are shown in Figure~\ref{fig:ew_vs_timescale}, and are characterised at integrated S/N ratios of $\sim2.0-4.4\sigma$ (N\,\textsc{iv}]), $\sim1.5-3.1\sigma$ (N\,\textsc{iii}]), $\sim2.4-8.9\sigma$ (C\,\textsc{iv}), and $\sim5.3-14.8\sigma$ (C\textsc{iii}]).


We find both C\,\textsc{iv} and C\,\textsc{iii}] are strongest in the bursty composites and weakest in the lulling ones at effectively every $\Delta t$, with differing EW(bursty)/EW(lulling) ratios according to both timescale and line. C\,\textsc{iii}] EWs remain approximately constant at all $\Delta t$, probing a small range of values $\sim3-4$ \AA\ (lulling) and $\sim7-9$ \AA\ (bursty), a factor of $\simeq 2.3$--$2.8\times$ increase between the two populations.
The difference in C\,\textsc{iv} EW between the bursty and lulling composites, however, varies according to timescale, with the largest contrast at $3$--$5$~Myr timescales ($\simeq 0.4$\,\AA\ for lulling composites and $5.3$--$5.5$\,\AA\ for bursty composites, a factor of $\simeq 14$) and the smallest at $10$--$20$~Myr timescales ($0.5$--$1.2$\,\AA\ for lulling composites and $2.4$--$2.9$\,\AA\ for bursty composites, falling from a factor of $\simeq 5$ to $\simeq 2$).
These trends reflect the different stars that power each line. C\,\textsc{iv} requires $>47.9$ eV photons, produced primarily by hot and massive O-type stars, thus making it a particularly apt tracer of intense and recent (a few Myr) massive star formation over $<10$\,Myr timescales, where the EW differences are largest, rather than longer $>10$\,Myr timescales where the EW differences reduce. This is consistent with \citet{robertsborsani2026}, who find that C\,\textsc{iv}-strong sources at $z > 10$ are powered by bursts of star formation on timescales $< 3$\,Myr.
C\textsc{iii}], on the other hand, requires only 24.4 eV and is thus powered by a broader range of longer-lived stellar populations, making it sensitive to star formation activity over both short and longer timescales. This has been shown by \citet{robertsborsani2024}, who demonstrate a ubiquity of C\,\textsc{iii}] emission in average prism spectra of emission-line sources at high redshift ($z>5$). The EWs seen here corroborate this interpretation, given the approximately constant trend with $\Delta t$.

The N\,\textsc{iii}] and N\,\textsc{iv}] equivalent widths are much weaker than their carbon counterparts, with $\simeq 0.2$--$1.0$\,\AA\ and $\simeq 0.6$--$1.5$\,\AA\ across $\Delta t = 3$--$20$\,Myr for each line, respectively.
Despite the weaker strengths, there are indications of a moderate trend. At $\sim3$--$5$\,Myr timescales, N\,\textsc{iii}] EWs are indistinguishable; at longer ($>5$\,Myr) timescales the bursty EWs become progressively stronger ($0.4$--$1.0$\,\AA) while the lulling EWs show no ordering with $\Delta t$, spanning $0.2$--$0.6$\,\AA. The difference between the two is most pronounced at $\Delta t=20$\,Myr, where the bursty EWs measure $1.0$\,\AA\ and the lulling EWs $0.6$\,\AA, a factor of $\simeq 1.5$, though the two remain consistent within their uncertainties.

In contrast, N\,\textsc{iv}] is strongest at the shortest timescales, spanning $0.9$--$1.5$\,\AA\ at $\Delta t=3$\,Myr and $0.6$--$1.1$\,\AA\ at $\Delta t=20$\,Myr. The decline is carried by the bursty composites, which fall from $1.5$\,\AA\ at $\Delta t=3$\,Myr, the highest of the three bins, to $0.6$\,\AA\ at $\Delta t=20$\,Myr, the lowest, while the lulling composites remain near $1.1$\,\AA\ throughout. This mirrors the behaviour of the C\,\textsc{iv} EWs, highlighting the analogous nature of the two lines and the extreme radiation fields that underpin them, though the N\,\textsc{iv}] contrasts remain within the uncertainties.

Overall, the values and trends shown here highlight both the prevalence and the utility of high-ionisation C and N lines as tracers of recent ($<100$\,Myr) star formation in $z>6$ galaxies. However, the observations also show them to be sensitive to different populations and timescales, with C\,\textsc{iv} and N\,\textsc{iv}] -- which require the hardest ionising photons -- to be particularly suitable tracers of short-lived, massive star formation over extremely short ($3-5$\,Myr) timescales; C\,\textsc{iii}] and N\,\textsc{iii}], on the other hand, trace a broader range of young stellar populations over a larger range of timescales.


\subsection{The Sensitivity of Abundance Ratios to Recent Star Formation}
\label{subsec:metalratios}
Having shown that the C\,\textsc{iii}], C\,\textsc{iv}, N\,\textsc{iii}] and N\,\textsc{iv}] line strengths respond to recent star formation, we now turn to the question of whether the chemical abundance ratios O/H, N/O and C/O are also sensitive to a galaxy's recent star formation history, and if so, whether they can be explained within standard chemical enrichment frameworks, or point to new ones.
We therefore measure the direct-$T_{\rm e}$ abundances of all twelve SFRE composites, defined in Section~\ref{sec:birthrate}, and examine how O/H, N/O and C/O vary with SFRE. The $\Delta t = 3$~Myr composites and the full-sample composite are summarised in Table~\ref{tab:stack_properties_b3}; the complete set of measured properties for every composite constructed in this work is given in Table~\ref{tab:stack_properties_all} of Appendix~\ref{app:allstacks}.

\begin{figure*}
    \centering
    \includegraphics[width=0.9\linewidth]{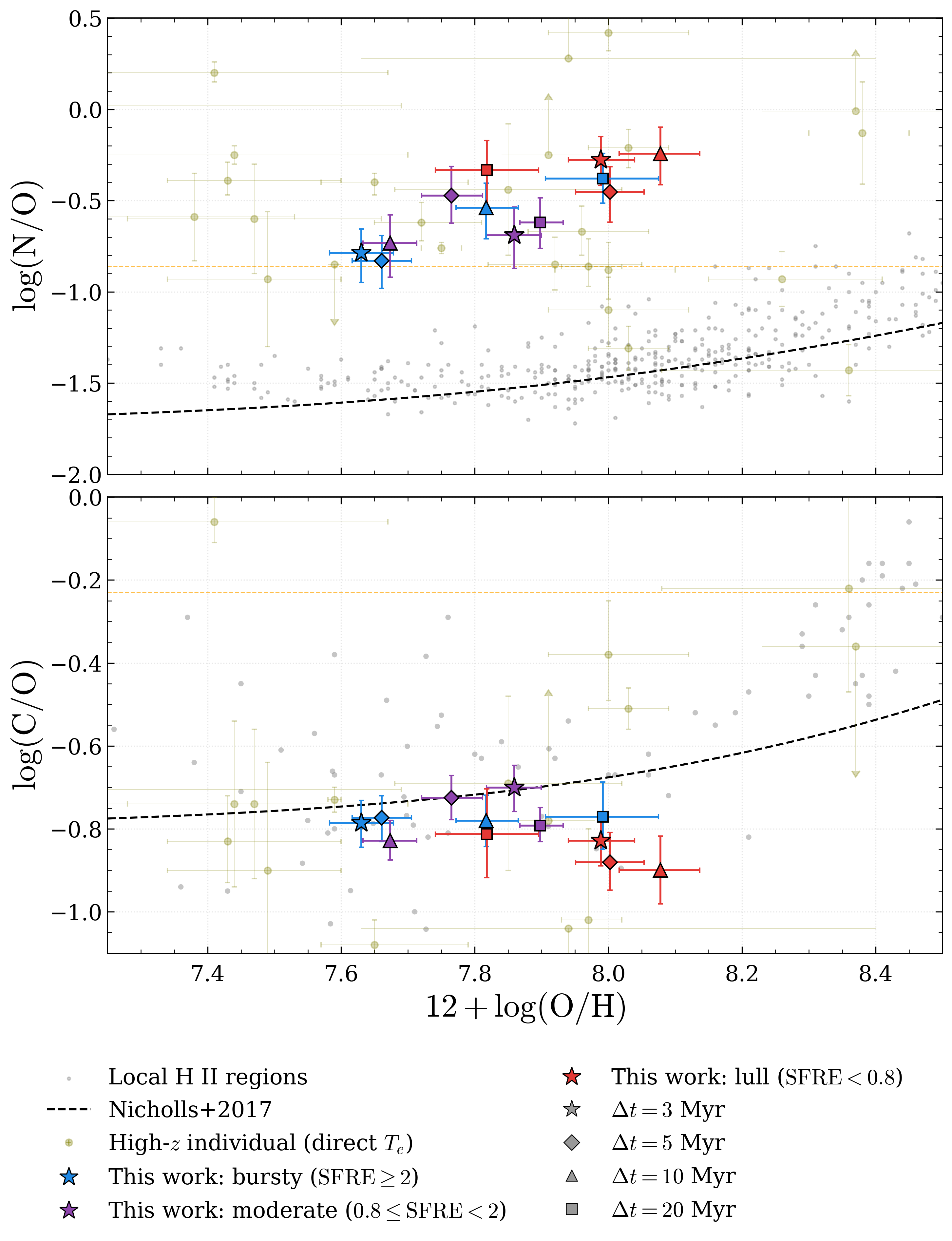}
    \caption{Nitrogen-to-oxygen (top) and carbon-to-oxygen (bottom) abundance ratios as a function of gas-phase oxygen abundance, derived from composite spectra binned by the star formation rate excess, SFRE$(\Delta t)$ (bursty in blue, moderate in purple, lulling in red), across various burst timescales ($\Delta t=$3, 5, 10, 20 Myr). Grey coloured points show local H\,\textsc{ii} region measurements, while the dashed black line shows the fit of \citet{nicholls2017}. Orange dashed lines mark solar values for N/O and C/O. Olive circles show individual high-redshift galaxies with direct $T_e$-based abundances from the literature \citep{james2009, berg2016, marques2024, vanzella2022, welch2025, stiavelli2025, labbe2024, zhang2025, arellanocordova2025, ubler2023, ji2024, tacchella2025, topping2024, isobe2023, topping2025, navarrocarrera2025, schaerer2024, curti2025, napolitano2025, cameron2023, calabro2024}, with arrows indicating upper limits.
    }
    \label{fig:birthrate_NOCO}
\end{figure*}

\begin{table*}
  \centering
  \renewcommand{\arraystretch}{1.5}
  \begin{tabular}{lcccc}
    \hline\hline
    Property & Full & Lull & Moderate & Bursty \\
    & $N=135$ & $N=70$ & $N=37$ & $N=26$ \\
    \hline
    \multicolumn{5}{c}{\textit{Sample properties}} \\ \hline
    Median redshift & $6.91^{+1.31}_{-0.64}$ & $6.82^{+1.13}_{-0.55}$ & $7.04^{+1.36}_{-0.77}$ & $7.21^{+0.96}_{-0.95}$ \\
    Median $M_{\rm UV}$ & $-19.73^{+0.49}_{-0.67}$ & $-19.82^{+0.46}_{-0.76}$ & $-19.70^{+0.39}_{-0.60}$ & $-19.58^{+0.50}_{-0.49}$ \\
    Median $\log(M_\star/M_\odot)$ & $8.65^{+0.44}_{-0.39}$ & $8.74^{+0.42}_{-0.38}$ & $8.72^{+0.25}_{-0.36}$ & $8.25^{+0.54}_{-0.43}$ \\
    \multicolumn{5}{c}{\textit{Chemical abundances \& conditions}} \\ \hline
    EW([O\,\textsc{iii}]$+$H$\beta$) (\AA) & $788^{+10}_{-10}$ & $559^{+9}_{-9}$ & $1022^{+19}_{-18}$ & $1219^{+29}_{-29}$ \\
    $12+\log(\mathrm{O/H})$ (direct) & $7.79^{+0.03}_{-0.03}$ & $7.99^{+0.05}_{-0.05}$ & $7.86^{+0.04}_{-0.04}$ & $7.63^{+0.05}_{-0.05}$ \\
    $12+\log(\mathrm{O/H})$ (strong) & $7.71^{+0.18}_{-0.15}$ & $7.89^{+0.25}_{-0.21}$ & $7.65^{+0.15}_{-0.13}$ & $7.59^{+0.13}_{-0.12}$ \\
    $\log(\mathrm{N/O})$ & $-0.57^{+0.09}_{-0.10}$ & $-0.28^{+0.12}_{-0.14}$ & $-0.69^{+0.15}_{-0.18}$ & $-0.79^{+0.13}_{-0.16}$ \\
    $\log(\mathrm{C/O})$ & $-0.80^{+0.03}_{-0.03}$ & $-0.83^{+0.07}_{-0.06}$ & $-0.70^{+0.06}_{-0.05}$ & $-0.79^{+0.06}_{-0.06}$ \\
    $\log(\mathrm{Ne/O})$ & $-0.78^{+0.01}_{-0.01}$ & $-0.74^{+0.02}_{-0.02}$ & $-0.78^{+0.02}_{-0.02}$ & $-0.78^{+0.03}_{-0.03}$ \\
    $A_V$ (mag) & $0.24^{+0.04}_{-0.04}$ & $0.18^{+0.06}_{-0.06}$ & $0.16^{+0.05}_{-0.05}$ & $0.26^{+0.08}_{-0.08}$ \\
    $\log U$ & $-2.97^{+0.01}_{-0.01}$ & $-3.27^{+0.02}_{-0.02}$ & $-3.03^{+0.02}_{-0.02}$ & $-3.00^{+0.02}_{-0.02}$ \\
    \hline
  \end{tabular}
    \caption{Physical properties of the $\Delta t = 3$~Myr SFRE-binned composite spectra and the full-sample composite ($6 \leq z \leq 10$, $-21<M_{\rm UV}<-19$). Abundances use the direct-$T_e$ method with the ICFs of \citet{martinez2025}. The abundances, $A_V$, $\log U$, EW, $T_e$ and $n_e$ are measured from the composite spectrum itself, whereas $z$, $M_{\rm UV}$ and $\log(M_\star/M_\odot)$ are the medians over the galaxies contributing to each composite. All quoted ranges are 16th--84th percentiles, giving the measurement uncertainty in the former case and the spread across the sample in the latter.}
  \label{tab:stack_properties_b3}
\end{table*}

The resulting measurements are plotted in Figure~\ref{fig:birthrate_NOCO}, which shows N/O and C/O as a function of overall metallicity O/H, for each burst timescale. Compared across the four burst timescales, the lulling composites are both the most metal-rich and the most nitrogen-enhanced, with a mean value $12+\log(\mathrm{O/H}) = 7.97$ (range $7.82$--$8.08$) and $\log(\mathrm{N/O}) = -0.33$ (range $-0.45$ to $-0.24$). The bursty composites lie systematically lower in both quantities, with a mean value $12+\log(\mathrm{O/H}) = 7.78$ (range $7.63$--$7.99$) and $\log(\mathrm{N/O}) = -0.64$ (range $-0.83$ to $-0.38$). The offset between bursty and lulling measurements is therefore $0.20$\,dex in O/H and $0.31$\,dex in N/O, and the two N/O ranges overlap over only $\simeq 0.07$\,dex compared with $\simeq 0.17$\,dex in O/H. C/O shows no comparable separation: the means are $\log(\mathrm{C/O}) = -0.78$ (bursty) and $-0.86$ (lulling), a difference of less than $0.1$\,dex, and all twelve composites span only $\simeq 0.20$\,dex with no significant ordering by $\mathrm{SFRE}$.

The burst timescale $\Delta t$ controls how cleanly the three SFRE bins separate. At the shortest timescales $\Delta t = 3$ and $5$\,Myr, the bursty composite is the most metal-poor and the least N/O-enhanced, and the abundances increase monotonically in both O/H and N/O as the SFRE shifts towards lulling systems. At longer timescales of $\Delta t = 10$\,Myr, this sequence is no longer monotonic, with the moderate composite falling below the bursty one in both the aforementioned abundance ratios. By $\Delta t = 20$\,Myr the O/H ordering has reversed altogether, with the bursty composite now the most metal-rich ($12+\log(\mathrm{O/H}) = 7.99^{+0.08}_{-0.09}$) and the lulling one the most metal-poor ($7.82^{+0.08}_{-0.08}$), although the two differ by only $0.17$\,dex ($\simeq 1.4\sigma$). The bursty--lulling contrast in N/O has correspondingly collapsed, from $0.51$\,dex at $\Delta t = 3$\,Myr to $0.05$\,dex ($-0.38$ against $-0.33$).
This degradation is partly methodological, since for $\Delta t \gtrsim 10$\,Myr $\mathrm{SFRE}(\Delta t)$ no longer isolates the present ionising population and the bins increasingly mix burst phases; the physical origin of this inversion is developed in Section~\ref{sec:kob_fer}.

The distinction and trend in abundance ratios are most evident in the 3\,Myr composites. From these, the direct-$T_e$ oxygen abundance rises from $12+\log(\mathrm{O/H}) = 7.63^{+0.05}_{-0.05}$ in the bursty bin to $7.86^{+0.04}_{-0.04}$ (moderate) and $7.99^{+0.05}_{-0.05}$ (lulling), a spread of $\simeq 0.36$\,dex, making the bursty composite the most metal-poor of the twelve $\mathrm{SFRE}$ stacks. N/O follows the same trend but more steeply, rising from $\log(\mathrm{N/O}) = -0.79^{+0.13}_{-0.16}$ in the bursty composite to $-0.69^{+0.15}_{-0.18}$ (moderate) and $-0.28^{+0.12}_{-0.14}$ (lulling), enriching the lulling stack by $\simeq 0.5$\,dex relative to the bursty one.
C/O, by contrast, is effectively flat within the uncertainties, with $-0.79^{+0.06}_{-0.05}$ (bursty), $-0.70^{+0.06}_{-0.05}$ (moderate), and $-0.83^{+0.07}_{-0.06}$ (lulling), spanning only $\simeq 0.13$\,dex. Galaxies caught in the strongest and most recent bursts are therefore systematically metal-poor and N/O-deficient relative to those in a post-burst lull. Because the contrast is clearest at the shortest timescales, this suggests the abundance patterns are likely driven by the present ionising population, rather than by the integrated star formation history of the source, and possibly affected by recent inflows and outflows of gas. We explore plausible scenarios in Section~\ref{sec:kob_fer}.

We perform three tests to verify that the abundance ratios between the lull and bursty composites are not driven by systematics. First, we measured the empirical per-pixel continuum S/N ratio of each stack from the median flux and standard deviation in narrow ($10$~\AA{}), line-free windows immediately flanking the N\,\textsc{iv}] $\lambda\lambda$1483,1486 and N\,\textsc{iii}] $\lambda\lambda$1749,1752 doublets. In our SFRE$(\Delta t = 3~\mathrm{Myr})$-stacks, the lull and moderate composites have both the highest N/O and the highest continuum SNR ($\sim 7\sigma$ per pixel), while the bursty composite has the lowest N/O and the lowest SNR ($\sim 4\sigma$ per pixel). A noise-driven enhancement of N/O would preferentially appear in the noisiest spectra; instead the highest N/O coincides with the highest-S/N continua, and the lowest N/O with the noisiest. We therefore conclude that the relative depths of the composites are unlikely to drive the differences in N/O seen here. Second, to verify that any incomplete accounting for dust is not driving the contrast in abundance ratios, we repeat the full analysis without accounting for dust obscuration and find near identical results: the qualitative pattern of Figure~\ref{fig:birthrate_NOCO} is unchanged, with the lull composites remaining the most metal-rich and the most N/O-elevated. Third, we verify that the metallicity ordering is not simply a reflection of small stellar mass differences between our composites. We adopt our fiducial $\mathrm{SFRE}(\Delta t = 3~\mathrm{Myr})$ composites -- where the contrast is greatest and clearest -- together with the $z = 4$--$10$ mass--metallicity relation (MZR) of \citet{nakajima2023}, and evaluate the predicted metallicity at the median stellar mass of each composite (Table~\ref{tab:stack_properties_b3}). The bursty composites are lower in stellar mass than the lull composites by $\simeq 0.4$\,dex ($\log(M_\star/M_\odot) = 8.25$ versus $8.74$), which on the MZR corresponds to a predicted metallicity difference of only $\simeq 0.1$\,dex ($12+\log(\mathrm{O/H}) = 7.77$ versus $7.87$). The offset we measure directly from the spectra is $0.36$\,dex, nearly four times larger than the MZR predicts, and therefore cannot be attributed to the mass--metallicity relation alone.

\section{Discussion}
\label{sec:discussion}

Our analysis yields two key results. First, nitrogen enrichment is ubiquitous across our emission-line-selected population at $6\leq z \leq10$, rather than being confined to a handful of extreme systems. This contrasts with prism-based studies, both at $z > 10$ \citep[e.g.][]{bunker2023gnz11, castellano2024} and at lower redshifts \citep[e.g.,][]{marques2024, zhu2026}, which recover elevated N/O only in isolated extreme systems, and highlights both the power and necessity of deep, $R\sim1000$ spectroscopy over statistical samples. Second, the resulting supersolar N/O and O/H abundance ratios depend systematically on recent star formation: at short timescales ($\Delta t \leq 5\,\mathrm{Myr}$), where the UV spectra are dominated by the strongest and most short-lived stellar populations, sources undergoing a relative lull in star formation activity are also the ones most enriched in both O/H and N/O, compared to their more intensely (or bursty) star-forming counterparts. In what follows we use these findings to identify the primary driver(s) of nitrogen enrichment at $6 \leq z \leq 10$, distinguishing possible delayed enrichment by AGB stars from more prompt channels (Wolf Rayet winds, supermassive stars and a top-heavy IMF; \citealt[e.g.,][]{berg2025, marqueschaves2026, cameron2024}), comparing our results to cosmological simulations, and assessing the implications for star formation and chemical enrichment in the earliest galaxies.


\subsection{The Prevalence of Nitrogen Enrichment at High Redshift}
\label{sec:prevalence}

The full-sample composite shown in Figure~\ref{fig:full_stack} is characterised by strong N\,\textsc{iv}] line emission, which combined with a direct-$T_{\rm e}$ oxygen abundance translates to supersolar N/O ($\log(\mathrm{N/O}) = -0.57^{+0.09}_{-0.10}$) 
and lies $\approx 1$~dex above local H\,\textsc{ii}-region values at comparable galaxy metallicity.
Since this supersolar N/O emerges from a high S/N median composite, such enhancement must represent a generic, population-wide property of our $6 \leq z \leq 10$ sample rather than a feature of a handful of extreme and luminous systems.

To place this ubiquity into context, we compare our findings with the literature across a number of redshift regimes. At $z \lesssim 4$, supersolar N/O represents a much rarer feature: among the $\simeq 50\,000$ star-forming galaxies at $z < 0.96$ with direct-$T_{\rm e}$ abundances in DESI DR2, \citet{scholte2026} identify only 24 ($\approx 0.05$ per cent) with elevated N/O at low metallicity. Similarly, at slightly higher redshifts of $z \sim 3$, \citet{schaerer2026} measure a large spread of subsolar values and a mean of $\log(\mathrm{N/O}) = -1.29$ in typical Lyman-break galaxies and Lyman-$\alpha$ emitters, statistically indistinguishable from $z \sim 0$ DESI samples at comparable metallicity. By contrast, at $4 < z < 10$, supersolar N/O becomes much more common, and our measurement is corroborated by several stacking analyses over comparable redshifts. \citet{tripodi2026} measure $\log(\mathrm{N/O}) \simeq -0.4$ in stacked prism spectra of $z > 4$ Ly$\alpha$ emitters, and \citet{umeda2026} similarly recover a supersolar N/O ($\log(\mathrm{N/O})_{\rm UV} = -0.20 \pm 0.24$), spanning a broad mass range down to $M_\star \sim 10^{5.7}\,M_\odot$, from the N\,\textsc{iv}]\,$\lambda\lambda1483,1486$ lines in their $R\sim1000$ stack at $z \simeq 4.5$--$10$, both consistent with our value of $\log(\mathrm{N/O}) = -0.57^{+0.09}_{-0.10}$. The prism stacks of \citet{hayes2024}, comprising $\simeq 1000$ galaxies at $z = 4$--$10$ binned by [O\,\textsc{iii}] equivalent width, yield a comparable enhancement with $\log(\mathrm{N/O}) = -0.42 \pm 0.70$ in their most nitrogen-enriched bin and falling to $-1.05 \pm 0.52$ in their least enriched, and both O/H and N/O rising as $W_{\mathrm{[O\,\textsc{iii}]}}$ decreases, mirroring the bursty-to-lull trend of our SFRE-binned stacks. These population-level measurements are also complemented by a growing number of individual $z > 4$ detections of nitrogen-enhanced systems \citep[e.g.,][]{topping2024, topping2025, senchyna2024, marques2024}. At $z > 10$, the most extreme individual N\,\textsc{iv}] and/or N\,\textsc{iii}] emitters (e.g. GN-z11 at $z = 10.6$; \citealt{bunker2023gnz11}, GHZ2/GLASS-z12 at $z = 12.3$; \citealt{castellano2024}, and the nitrogen-emitting MoM-z14 at $z = 14.4$; \citealt{naidu2026}) reach even higher nitrogen-enrichment and are attributed to prompt channels.

Although there is likely a degree of overlap between some of the population-wide data sets discussed here -- and thus some underlying agreement -- this analysis builds on previous works by affording a unique combination of a large sample size characteristic of the luminous population, spectral resolution and unprecedented depth, and precision abundance measurements with accurate ionisation frameworks, to allow us to place some of the most robust and representative constraints at high redshift thus far.
To this end, comparing our results to those at similar and/or lower redshift, enhanced N/O appears to grow more prevalent with increasing redshift: extremely rare in the local Universe and moderately rare at intermediate redshift, before becoming a common feature of the star-forming population at $6 \leq z \leq 10$. This evolution highlights that star formation and chemical enrichment proceed differently at $z \sim 0$ than in the first billion years.

\subsection{Delayed AGB Enrichment, CCSNe, and Inflows as Regulators of the Abundance Patterns}
\label{sec:kob_fer}

Having measured the direct-$T_e$ abundances of the full-sample composite and of the SFRE-binned stacks, and established that N/O is ubiquitous, supersolar and peaks in the lull composites on short timescales, we now turn to interpreting their physical origin.

We interpret our measurements within the dual-starburst galactic chemical evolution (GCE) model of \citet{kobayashi2024}, which has been compared to a growing number of high-redshift nitrogen emitters, from the extreme individual emitter GN-z11 \citep{kobayashi2024} to larger samples of nitrogen-enhanced galaxies \citep[e.g.,][]{bhattacharya2026, rusakov2026}. In this model the extreme N/O ratios of high-$z$ galaxies are reached through an intermittent star formation history: a sequence of bursts separated by periods of relative quiescence, in which each burst is triggered by a pristine gas inflow that dilutes the interstellar medium and lowers O/H, and is followed by a galactic outflow that temporarily reduces or halts star formation and vents enriched gas before the next cycle begins. Following \citet{kobayashi2024}, the assumption is that stellar ejecta mix instantaneously into the interstellar medium once released. The release itself, however, occurs over a variety of timescales depending on the origin considered, whether these be stellar winds of Wolf Rayet and/or very massive stars ($\lesssim$ a few Myr; \citealt{schaerer1998, crowther2007}), CCSNe ($\sim 3$--$50$~Myr; \citealt{zapartas2017}), and/or delayed return from intermediate-mass AGB stars ($\sim 100$--$300$~Myr; \citealt{henry2000, karakas2010}).

Pristine gas inflows are invoked in this model as the fuel and trigger of each burst of star formation \citep[e.g.,][]{kobayashi2024, rizzuti2025, bhattacharya2026, pascalau26}, in line with cosmological simulations that predict cold, metal-poor streams from the cosmic web \citep{dekel2009}. Such accretion can sustain star formation while also diluting the gas-phase metallicity \citep{sanchezalmeida2014}, and has been inferred observationally from inverse metallicity gradients at $z \sim 3$ \citep{cresci2010} as well as from offsets to the fundamental metallicity relation at $z \sim 8$ \citep{heintz2023}. Metal dilution can be probed directly by our direct-$T_{\rm e}$ O/H measurements across the composites. At the shortest timescales the bursty composites are the most oxygen-poor of the stacks, with O/H rising monotonically as the composites approach a lulling phase: at $\Delta t = 3$~Myr from $12+\log(\mathrm{O/H}) = 7.63 \pm 0.05$ (bursty), through $7.86$ (moderate), to $7.99 \pm 0.05$ (lull), a span of $0.36$~dex, with the same ordering at $\Delta t = 5$~Myr ($7.66 \to 7.77 \to 8.00$). This behaviour is expected if the same pristine inflow that supplies the fuel for eventual star formation temporarily dilutes the oxygen-rich ISM, and importantly this would not impact the measured N/O or C/O abundance ratios.

After a burst of star formation, nitrogen enrichment can also occur either from the release of previously enriched gas from earlier stellar generations (e.g., AGB stars), and/or from prompt channels (such as Wolf Rayet stars undergoing a nitrogen-rich phase, or very-to-super massive stars), both of which can increase N/O by up to an order of magnitude. In both cases, CCSNe resulting from the burst of star formation eventually produce and release oxygen into the ISM, thereby reducing the N/O ratio and increasing the overall metallicity. Although the most massive stars explode within a few Myr of the burst, they are rare and thus most of the oxygen comes from the more numerous $\sim 15$--$25\,M_\odot$ progenitors \citep{kroupa2001, limongichieffi2018}, whose $\gtrsim 10$~Myr lifetimes \citep{leitherer1999} set the timescale for CCSNe oxygen return. The effect of this timescale is seen in our composite measurements, where we find O/H values associated with the bursty stacks remain nearly constant from $\Delta t = 3$ to $5$~Myr ($7.63$, $7.66$), but rise to $7.82$ at timescales of $\Delta t = 10$~Myr (when CCSNe are expected to begin enriching the surrounding gas) and reach $7.99$ at $\Delta t = 20$~Myr. Eventually the burst subsides into a relative lull, during which the oxygen return from its CCSNe runs its course and ceases, while nitrogen continues to be supplied by the AGB descendants of earlier generations. N/O therefore recovers to its highest values at an elevated O/H, the configuration sampled by our lulling composites, until a subsequent pristine inflow dilutes the gas and triggers the next burst. 

\begin{figure}
    \centering
    \includegraphics[width=1\linewidth]{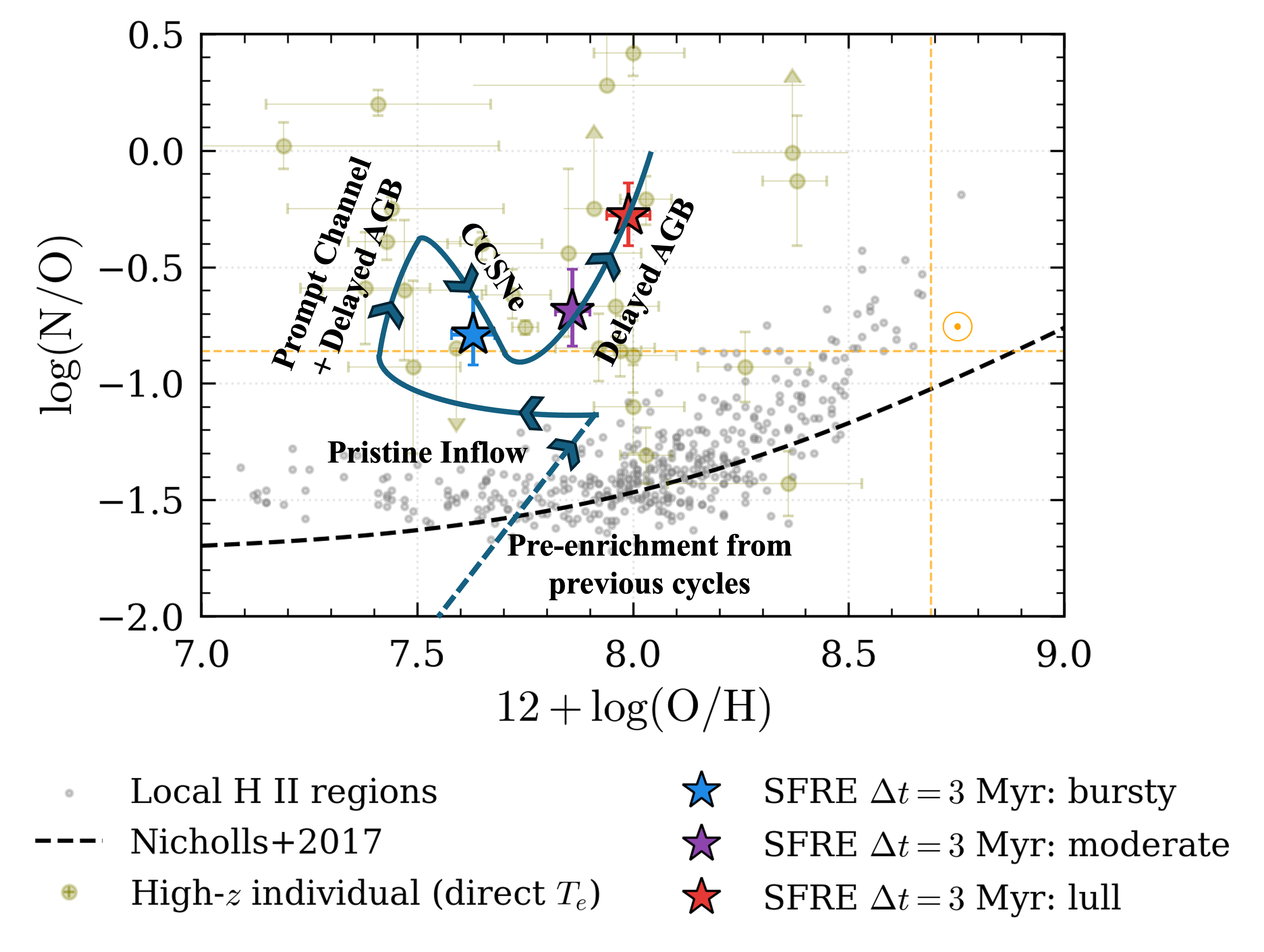}
    \caption{Schematic illustration of the burst-cycle interpretation developed in Section~\ref{sec:kob_fer}, shown in the $\log(\mathrm{N/O})$ versus $12+\log(\mathrm{O/H})$ plane. The curve is schematic and illustrates the ordering of the processes involved rather than a path followed by any individual galaxy. Coloured stars mark our $\Delta t = 3$~Myr star formation rate excess (SFRE) composites: bursty (blue), moderate (purple) and lull (red). Background points show the comparison samples, local H\,\textsc{ii} regions (grey) and high-$z$ individual direct-$T_e$ measurements (olive), with the \citet{nicholls2017} $z \sim 0$ relation (black dashed) and the solar values (orange dashed lines and $\odot$) shown for reference. The wavy blue curve traces the picture described in this section, based on the GCE models of \citet{kobayashi2024}. The cycle proceeds as follows: assuming an already nitrogen-enriched floor built up from preceding star formation (dashed arrow, ``Pre-enrichment''), a pristine gas inflow dilutes the gas and lowers O/H (``Inflow''), placing the galaxy in a metal-poor, bursty regime. AGB stars born in a previous burst, together with a possible contribution from the prompt winds of Wolf Rayet stars in some galaxies, then raise N/O within a few Myr of the burst. The burst's CCSNe subsequently raise O/H and lower N/O through oxygen production, and finally the delayed winds of an AGB population drive N/O back up.}
    \label{fig:interpretation}
\end{figure}

We emphasise that each composite is a median over a distinct subsample of galaxies rather than a single system followed through a cycle. The bursty--lulling contrast is therefore a population offset, combining the phase of the burst cycle with the differing enrichment histories of the galaxies occupying each phase, and its amplitude need not correspond to the excursion made by any individual galaxy. The scenario is summarised in the schematic in Figure~\ref{fig:interpretation}.

In contrast, the C/O ratios associated with our composites remain comparatively steady, without significant evolution with SFRE or burst timescale. For our fiducial $\Delta t = 3$~Myr composite, we measure $\log(\mathrm{C/O}) = -0.79^{+0.06}_{-0.05}$ (bursty), $-0.70^{+0.06}_{-0.05}$ (moderate) and $-0.83^{+0.07}_{-0.06}$ (lull), consistent with each other within uncertainties. A similar conclusion can be derived from comparisons across the full sample, which show $\log(\mathrm{C/O})$ ranges of $-0.79$ to $-0.77$ (bursty; mean of $-0.78$), $-0.83$ to $-0.70$ (moderate; mean of $-0.76$), and $-0.90$ to $-0.81$ (lulling; mean of $-0.86$), consistent with or moderately below the local relation \citep{nicholls2017,berg2016, berg2019}. The modest ranges and lack of clear evolution with SFRE, burst timescale, or metallicity are consistent with carbon delivery on short timescales from the same CCSNe that deliver oxygen \citep{limongichieffi2018}, rather than $\sim 100$--$300$~Myr delayed contributions from AGB stars. In the case of the latter, hot-bottom burning within the star converts nearly all of the freshly dredged-up carbon into nitrogen, before the envelope is expelled into the ISM \citep{karakas2010}.

The relative N contributions from AGB winds compared to Wolf Rayet or VMS/SMS enrichment are challenging to disentangle based on individual abundance ratios alone. A number of considerations point towards our galaxy sample being dominated by AGB winds, rather than WR or VMS/SMS, however. Firstly, the ubiquity of N\,\textsc{iv}] and supersolar N/O in both our full-sample composite and the composites discussed in Section~\ref{subsec:metalratios}, respectively, suggests an underlying floor of nitrogen enrichment that is challenging to explain with fast-acting channels alone, given their extremely brief duration and comparative rarity. For example, empirical estimates of WR populations suggest a fraction relative to O-type stars of $\sim0.15$  per cent in the solar neighbourhood (although there is some uncertainty based on the assumption of single- or binary-star systems; \citealt{eldridge2017}), which decreases significantly to $\sim0.02$ per cent at even lower (sub-SMC) metallicities \citep{crowther2007}. Moreover, the nitrogen-rich phase required for N/O enrichment typically lasts only $\sim0.03-0.3$\,Myr \citep{limongichieffi2018,berg2025}. Similarly, VMS are directly detected only in the densest local star clusters \citep{crowther2010,martins23}, and predicted nitrogen yields over $\lesssim2-3$\,Myr depend significantly on assumed IMF upper mass limits and cluster densities (e.g., \citealt{marqueschaves2024,charbonnel2023}). Secondly, the clear patterns of N/O and C/O with burstiness are inconsistent with expectations from dominant WR/VMS origins. In particular, these exotic origins predict the highest N/O ratios in the burstiest systems, which goes contrary to our observations where bursty systems trace the lowest N/O values. Moreover, the invariance of C/O against a $\sim0.5$ dex range in N/O is also inconsistent with the scatter expected from short-lived origins \textit{on a population level}, which vary significantly in their C, N, and O yields with time. Nonetheless, such origins could plausibly (temporarily) dominate individual systems or provide more minor contributions to the larger population, as has been seen through direct observational signatures in a number of nitrogen-enhanced $z>6$ objects \citep{berg2025,marqueschaves2026}.

We nevertheless search our composites for direct spectral signatures of WR stars or VMS, namely broad He\,\textsc{ii}\,$\lambda1640$ in the rest-frame UV and the optical ``blue bump'' near $4650$\,\AA\ \citep{schaerer1998, brinchmann2008, berg2025} for the former, and a strong N\,\textsc{v}\,$\lambda 1240$ P-Cygni profile \citep{marqueschaves2026} for the latter. None of our composites, however, reveal signatures of a WR bump, although all but the full-sample composite may lack the S/N to rule out such weak emission. The full-sample composite does instead show noticeable broad absorption blueward of N\,\textsc{v}, plausibly indicating stellar winds from VMS. As discussed in Section~\ref{sec:full_stack}, however, modelling of the feature -- along with the spectrum's observed C\,\textsc{iv} P-Cygni feature -- requires no exotic stellar origin and is instead well reproduced by standard SSP templates that include binary-stripped stars and the radiatively driven winds of massive OB populations. As such, our observations point to regulation of abundance ratios primarily by pristine gas inflows, CCSNe, and AGB production, with more minor contributions from WR stars. Considering the supersolar floor probed here, our sample of $6 \leq z \leq 10$ sources likely trace systems which have undergone continuous nitrogen enrichment from multiple cycles of AGB star formation and enrichment.

\subsection{Comparison to Cosmological Simulations}

\begin{figure*}
    \centering
    \includegraphics[width=1\linewidth]{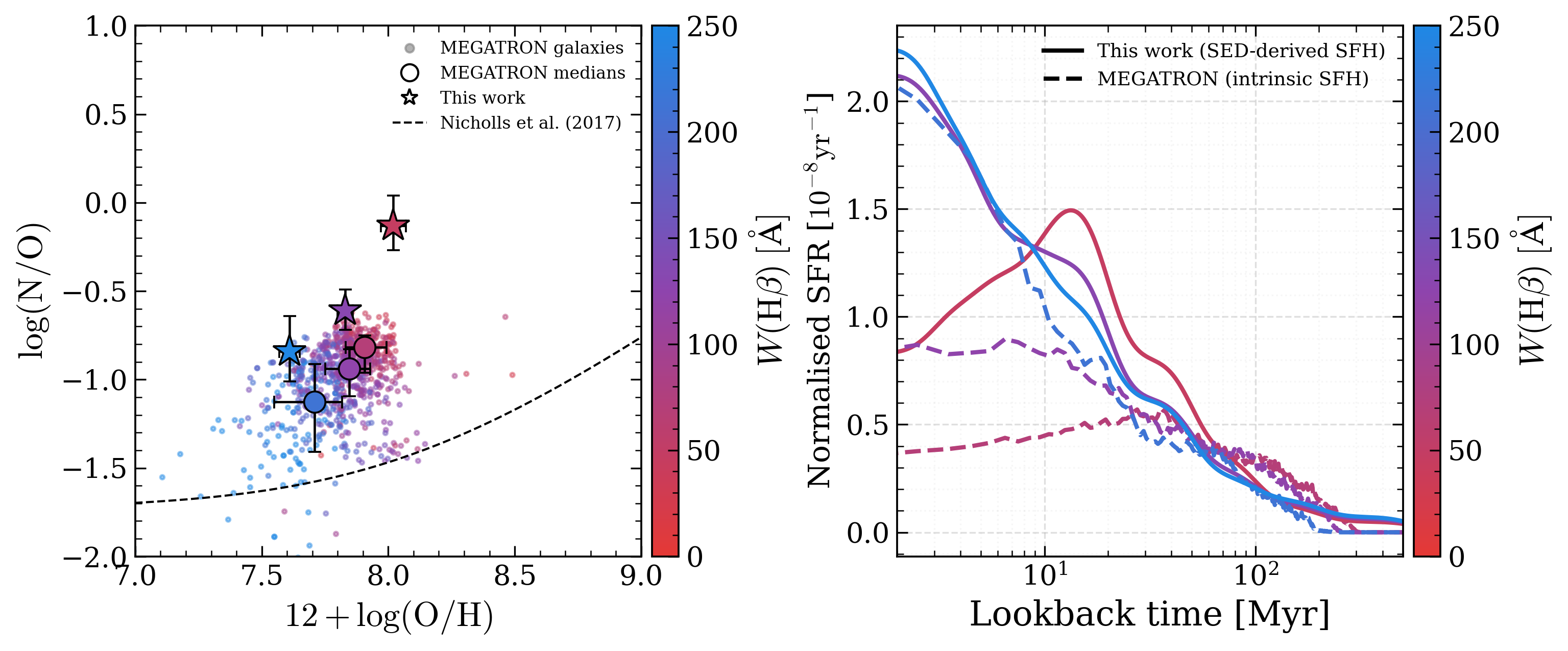}
    \caption{Comparison of the \textsc{MEGATRON} ``efficient star formation'' model with our NIRSpec data, binned by $W_{\mathrm{H}\beta}$. \textit{Left:} Direct-method $\log(\mathrm{N}/\mathrm{O})$ against $12+\log(\mathrm{O}/\mathrm{H})$. Small points are individual \textsc{MEGATRON} galaxies, with abundances measured from the integrated spectrum within $0.25\,R_{\rm vir}$ using the same direct-$T_e$ pipeline applied to our data, coloured by H$\beta$ equivalent width. Circles and stars indicate the \textsc{MEGATRON} and NIRSpec composites, respectively, in three identical $W_{\mathrm{H}\beta}$ bins. \textit{Right:} Median stellar-mass-normalised star formation histories of the same $W_{\mathrm{H}\beta}$ bins, coloured as in the left panel. Solid curves show our SED-derived stacked SFHs and dashed curves the \textsc{MEGATRON} intrinsic SFHs. Uncertainties in SFHs are not included in this plot for clarity.}
    \label{fig:megatron-NO}
\end{figure*}

We now compare the abundance measurements from our composites with the \textsc{MEGATRON} cosmological simulations \citep{katz2025, choustikov2026}, a suite of high-resolution zoom-in models allowing us to follow the chemical enrichment patterns of galaxies in the early Universe ($z > 8.5$). An advantage of \textsc{MEGATRON} is the prediction of emission-line spectra rather than intrinsic abundances alone, which can be analysed in the exact same fashion as our real spectra (see Section~\ref{sec:chemical_abund}), in order to extract observed abundances with which a direct comparison can be made. Any difference between the observed and simulated data then reflects the limitations of the underlying chemical enrichment models rather than systematics from the way the abundances were measured. The \textsc{MEGATRON} suite contains four runs that differ in their star formation and feedback prescriptions, and we use the fiducial ``efficient star formation'' (ESF) model. Because this run retains rather than expels its enriched gas, it builds the densest and most metal-rich ISM of the four, and the highest mass--metallicity normalisation \citep{choustikov2026}. Its nitrogen is produced solely by CCSNe and AGB stars under a fixed IMF; the simulations do not model Wolf Rayet stars, and include no top-heavy or variable IMF, super-AGB, or supermassive-star channel. It is therefore a test of whether CCSNe and AGB enrichment alone, in a fully resolved multi-phase ISM, can reproduce the trends we observe.

To build comparable measurements, we select every galaxy in the \textsc{MEGATRON} ESF run, with a similar UV-magnitude range to our spectroscopic sample (Section~\ref{sec:sample}); since the $-21 < M_{\rm UV} < -19$ cut of our observational sample leaves too few simulated galaxies for robust statistics, we widen the selection to $-22 < M_{\rm UV} < -18$, and note that this does not affect the median properties of the simulated sample. For each source, we then sum the emission from all gas cells within $0.25\,R_{\rm vir}$ (the aperture used by \citet{choustikov2026} to capture the full ISM of the central galaxy) to form an integrated spectrum analogous to our composites and to the integrated light JWST observes. We then measure O/H and N/O from the spectrum with the same direct-$T_e$ pipeline and \citet{martinez2025} prescriptions used for the data (Section~\ref{sec:chemical_abund}).

We bin both the simulated and our observed galaxies by their rest-frame H$\beta$ equivalent width, $W_{\mathrm{H}\beta}$; this is the same spectroscopic burstiness indicator we validate against the SFRE parameter in Section~\ref{sec:birthrate} and
has the added advantage of being measured directly from the spectrum, and thus can be applied in identical fashion to both samples. We stack both samples in three bins of $W_{\mathrm{H}\beta}$ ($W_{\mathrm{H}\beta} < 100$~\AA, $100 \leq W_{\mathrm{H}\beta} < 150$~\AA, and $W_{\mathrm{H}\beta} \geq 150$~\AA); four observed galaxies whose spectra are too noisy for the H$\beta$ line fit to converge are excluded, leaving 131 of the 135.

We plot the resulting abundance measurements and star formation histories for the $W_{\mathrm{H}\beta}$-binned composites in Figure~\ref{fig:megatron-NO}, where we find both the observed and simulated stacks recover the same N/O--O/H trends as the SFRE-binned composites in Figure~\ref{fig:birthrate_NOCO}. As shown in the left panel of Figure~\ref{fig:megatron-NO}, the \textsc{MEGATRON} galaxies follow the same $W_{\mathrm{H}\beta}$ trend as the data: the high-$W_{\mathrm{H}\beta}$ galaxies, caught after a recent burst, are more metal-poor and less N/O-enhanced, whereas the low-$W_{\mathrm{H}\beta}$ galaxies, in a relative lull, are more metal-rich and more N/O-enhanced. The median oxygen abundance rises from $12+\log(\mathrm{O/H}) = 7.71$ in the bursty (high-$W_{\mathrm{H}\beta}$) bin to $7.91$ in the lull (low-$W_{\mathrm{H}\beta}$) bin, and $\log(\mathrm{N/O})$ from $-1.13$ to $-0.82$. The simulation therefore reproduces the directional trend of increasing metallicity and N/O from bursty sources to lulling sources with standard CCSNe and AGB enrichment alone.

The right panel of Figure~\ref{fig:megatron-NO} shows the corresponding stellar mass-normalised star formation histories, where those of the real data are measured as in Section~\ref{sec:sed_fitting} and the simulated ones are the intrinsic \textsc{MEGATRON} ones. In both the simulation and our data, the three $W_{\mathrm{H}\beta}$ bins differ only in their recent star formation: the high-$W_{\mathrm{H}\beta}$ bin is rising strongly by the epoch of observation, the intermediate bin shows mildly elevated recent activity, and the low-$W_{\mathrm{H}\beta}$ bin shows declining activity, with its peak having occurred $\sim 10$--$15$~Myr earlier. Beyond $\sim 20$--$30$~Myr of lookback time the mean histories of the three bins converge and become effectively indistinguishable.
Measurements of SFRE from these SFHs corroborate the observations, with the high-$W_{\mathrm{H}\beta}$ bin returning the largest values on the shortest windows and the low-$W_{\mathrm{H}\beta}$ bin the smallest, and the separation diminishing as $\Delta t$ approaches that convergence timescale.

Despite the agreement in N/O versus metallicity and recent SFH, we note that at fixed metallicity the \textsc{MEGATRON} galaxies sit systematically below our stacks in N/O, by $\approx 0.3$--$0.7$\,dex across all bins. While some minor differences are to be expected
such an offset may indicate 
missing nitrogen enrichment in the simulated galaxies.

Two candidate sources are absent from \textsc{MEGATRON} by construction. The first is Wolf Rayet enrichment. In principle the nitrogen deposited by WR winds could be retained in the interstellar medium and accumulate over successive bursts, and chemical evolution models of RXCJ2248-ID3, a $z \simeq 6.1$ galaxy with a confirmed Wolf Rayet population, show that WR winds can raise N/O by $\gtrsim 0.5$~dex \citep{berg2025}. Whether this scales to the population average is, however, uncertain: WR stars descend from O-type progenitors with $M_\star \gtrsim 20$--$25\,M_\odot$, which under a standard IMF constitute $\lesssim 1$ per cent of each stellar generation, and the nitrogen-rich WN phase is brief and followed by carbon- and oxygen-enriched ejecta from the same stars, which act to dilute the transient N/O enhancement. A cumulative WR contribution is therefore plausible for individual systems such as RXCJ2248-ID3, but is unlikely, on its own, to supply the population-wide offset we measure. The second, and less constrained, candidate is the super-AGB channel, whose massive ($\sim 8$--$10\,M_\odot$) progenitors undergo strong hot-bottom burning and return nitrogen within a few tens of Myr. This return time is short enough for the channel to operate within the burst cycles our composites probe, so any super-AGB contribution would be present in the measured N/O; the simulation, however, does not include these yields. Either contribution would act on the absolute normalisation of the relation rather than on the $W_{\mathrm{H}\beta}$ gradient itself, which both the data and the simulation already reproduce, but the balance between them, and whether they suffice, remains open.


The \textsc{THESAN-ZOOM} suite \citep{kannan2025} provides a second set of high-resolution cosmological zoom-in simulations, run down to $z = 3$. Unlike the \textsc{MEGATRON} simulation, the \textsc{THESAN-ZOOM} suite tracks the elemental abundances of the gas directly rather than predicting the emission-line fluxes, so we compare only against its predicted N/O and O/H. The analysis of \citet{mcclymont2025} uses this suite to study how bursty star formation drives the chemical evolution of these galaxies, finding that it generates order-of-magnitude variations in gas-phase N/O within $\lesssim 100$~Myr from CCSNe and AGB enrichment. In this scenario, CCSNe feedback preferentially expels the oxygen-enriched gas, and the stars formed in the burst remain behind; the nitrogen-rich winds of their AGB descendants then enrich the depleted ISM, whose reduced mass allows even modest AGB yields to raise N/O substantially, a process repeated over successive cycles of bursty star formation. \citet{mcclymont2025} further find that nitrogen-rich galaxies in their simulation typically lie below the star-forming main sequence, consistent with our picture in which the lull composites sample the AGB-enriched gas reservoir before the next burst of star formation re-enriches the gas with oxygen from CCSNe. Qualitatively, our results align with the \textsc{THESAN-ZOOM} interpretation; the incorporation of feedback yields one marked difference: since oxygen is expelled by outflows, \citet{mcclymont2025} show that lulling galaxies attain high N/O at low O/H, whereas our composites indicate high N/O at high O/H. The burstiness framework thus accounts for the elevated N/O but does not reproduce the concurrent rise in O/H along the same sequence.
  

Both the \textsc{MEGATRON} and \textsc{THESAN-ZOOM} simulations reproduce a burst-to-lull modulation of N/O using only standard CCSNe and AGB enrichment, assuming standard stellar yields. Both show remarkable agreement with our spectroscopic measurements, reproducing the elemental abundance ratio gradients we observe as a function of recent star formation activity, although neither reproduce the observations completely, suggesting both feedback and minor nitrogen contributions from fast-acting channels may complicate the picture. Nonetheless, the significant agreement of our data with AGB- and CCSNe-based simulations supports the picture described in Section~\ref{sec:kob_fer} and corroborate delayed yields as the primary origin for nitrogen enrichment at $6 \leq z \leq 10$.


\section{Summary and Conclusions}

We have presented a stacking analysis of 135 emission-line-selected star-forming galaxies at $6 \leq z \leq 10$, observed with JWST/NIRSpec $R \sim 1000$ spectroscopy. By constructing high S/N composite spectra, we recover faint UV nitrogen, carbon and oxygen diagnostics that fall below the detection threshold of individual spectra. With these composites, we investigate the prevalence of high-ionisation line emission and the role of bursty star formation in regulating N/O, C/O, and O/H abundance ratios. Our key results are as follows:

\begin{enumerate}

    \item A median composite spectrum of the emission-line-selected full sample reveals high-ionisation (N\,\textsc{iv}], C\,\textsc{iv}, C\,\textsc{iii}], He\,\textsc{ii}\,$\lambda$1640) and auroral ([O\,\textsc{iii}]\,$\lambda$4363) line emission that is generally inaccessible in individual spectra at these redshifts. The presence of such lines, indicative of very hard radiation fields and young, massive stellar populations, suggests they are prevalent among the average $6\leq z \leq 10$ source. Making use of temperature- and density-sensitive lines within a multi-ionisation-phase framework, the lines yield supersolar $\log(\mathrm{N/O}) = -0.57^{+0.09}_{-0.10}$ alongside subsolar $\log(\mathrm{C/O}) = -0.80 \pm 0.03$ at a metallicity of $12+\log(\mathrm{O/H}) = 7.79 \pm 0.03$, suggesting \textbf{nitrogen enrichment is ubiquitous at these redshifts rather than a peculiarity of extreme sources.}

    \vspace{0.2cm}
    
    \item Resonant absorption is seen in both high- and low-ionisation phases, with low covering fractions ($C_f \simeq 0.3$--$0.5$, sample-averaged lower limits) indicating neither covers the continuum fully. The high-ionisation gas is also blueshifted relative to systemic by $v_\mathrm{cen} = -155 \pm 33$\,km\,s$^{-1}$, while the low-ionisation gas sits close to the systemic velocity on average ($v_\mathrm{cen} = -1 \pm 22$\,km\,s$^{-1}$). This combination of measurements indicates \textbf{a shift toward patchy interstellar medium gas, as well as the presence of multi-phase ionised outflows.}

    \vspace{0.2cm}

    \item We investigate the sensitivity of N\,\textsc{iv}], N\,\textsc{iii}], C\,\textsc{iv}, and C\,\textsc{iii}] equivalent widths and abundance ratios to recent star formation activity. By stacking according to recent star formation activity and burst timescale ($\Delta t = 3$, $5$, $10$, $20$), we find clear patterns linked to periods of relative ``bursty'', ``moderate'', or ``lulling'' activity. C\,\textsc{iii}] and C\,\textsc{iv} display the strongest EWs in bursty composites ($7.2$--$9.5$\,\AA\ and $2.4$--$5.5$\,\AA, respectively) and the weakest EWs in their lulling counterparts ($3.0$--$3.9$\,\AA\ and $0.4$--$1.2$\,\AA, respectively), with C\,\textsc{iv} peaking at $3$--$5$~Myr timescales (with a burst-to-lull ratio of $\simeq 14\times$). The nitrogen lines show comparatively weaker strengths, with N\,\textsc{iv}] appearing strongest in the burstiest systems with $1.5$\,\AA\ at $3$\,Myr timescales -- linking it and C\,\textsc{iv} to the most massive and shortest-lived stellar populations -- and progressively declining down to $0.6$\,\AA\ at $20$\,Myr, along with more moderate evolution in lulling systems. N\,\textsc{iii}] is weaker still ($0.2-1.0$\,\AA), with only a moderate increase in strength at longer timescales and with star formation activity. Comparing abundance ratios, on average lulling galaxies are characterised by the highest metallicities, $12+\log(\mathrm{O/H}) = 7.97$ and strongly supersolar $\log(\mathrm{N/O}) = -0.33$, whereas the burstiest galaxies reveal the lowest metallicities $12+\log(\mathrm{O/H}) = 7.78$ and $\log(\mathrm{N/O}) = -0.64$. The trend is most evident on the shortest timescales, $\Delta t = 3$~Myr, where N/O displays the most significant evolution ($\sim0.5$ dex). C/O ratios are found to be subsolar and nearly invariant around the local relation in all composites. \textbf{High-ionisation N\,\textsc{iv}] and C\,\textsc{iv} therefore track intense and recent star formation activity on the shortest burst timescales, while the N/O and O/H abundance ratios vary systematically with burst phase, pointing to well-defined enrichment origins.}

    \vspace{0.2cm}

    \item We compare the clear N/O, C/O and O/H patterns to theoretical frameworks, and interpret these as snapshots of galaxies caught at different stages of a duty cycle. In this scenario, pristine gas inflow first dilutes oxygen-rich gas and triggers star formation, then prompt oxygen production by CCSNe drives the N/O ratio down and increases overall metallicity, followed by Nitrogen-enhancement through delayed enrichment by AGB stars once star formation activity decreases, thereby raising N/O as a relative lulling phase begins. Prompt channels may momentarily boost N/O immediately following a burst, but are unlikely to account for the population-wide nitrogen excess when considering the near-constancy and small scatter of C/O, together with the lack of direct spectral signatures of WR stars and/or VMS. Repeated over successive bursts, \textbf{the abundance patterns and supersolar N/O floor are therefore consistent with snapshots of a recurring burst cycle, with AGB winds as its dominant source of nitrogen enrichment.}

    \vspace{0.2cm}

    \item Comparing to simulations, both the \textsc{MEGATRON} and \textsc{THESAN-ZOOM} suites generate the same burst-to-lull modulation of N/O using only CCSNe and AGB yields. In particular, direct comparisons to \textsc{MEGATRON} composite spectra establish the same links as the NIRSpec composites to bursty star formation over short timescales.
    Neither suite reproduces our measurements in full, however: \textsc{MEGATRON} sits $\approx 0.3$--$0.7$\,dex below our composites in N/O at fixed metallicity, while \textsc{THESAN-ZOOM} attains the elevated N/O without the accompanying rise in O/H. The agreement with these simulations therefore indicates that \textbf{population-wide nitrogen enhancement can be explained by a generic interplay of bursty star formation, gas flows and delayed AGB enrichment, although some prompt nitrogen production from WR stars may explain some minor differences.}
    
\end{enumerate}

This analysis demonstrates the power of stacking procedures combined with $R\sim1000$ spectroscopy to unveil the faintest and most sensitive probes of the ISM and stellar conditions dominating the earliest systems. By bypassing the depths, sample sizes, and physical biases that limit typical observations, this study transforms individual case studies into a precise and statistical mapping of early chemical enrichment, revealing how and when the first galaxies synthesised their heavy elements. Extending this approach to the characteristically fainter and burstier population below our luminosity range, within reach of lensing cluster fields, represents the next step and will test whether star formation and chemical synthesis proceed similarly across the entire galaxy population, or require unexplored origins.

\section*{Acknowledgements}
RSR acknowledges support from the Science and Technology Facilities Council (STFC) through a PhD studentship. The computationally intensive elements of this work (data reduction and SED fitting) were performed using the UCL Myriad High Performance Computing Facility (Myriad@UCL) and associated support services. The authors thank Nicholas Choustikov for providing emission-line measurements from the \textsc{MEGATRON} simulations \citep{choustikov2026} and for valuable discussions on their workings and comparison with our results; Pietro Bergamini for providing magnification factors for galaxies observed in the Abell 2744 cluster \citep{abell2744lense}; Richard Brooks for valuable discussions on the burst cycle; and Jason Sanders for valuable discussions and feedback on the manuscript. This work is based on observations made with the NASA/ESA/CSA James Webb Space Telescope, obtained from the Mikulski Archive for Space Telescopes at the Space Telescope Science Institute, which is operated by the Association of Universities for Research in Astronomy, Inc., under NASA contract NAS 5-03127 for JWST.

\section*{Data Availability}
 
The JWST/NIRSpec data underlying this article are publicly available from the Mikulski Archive for Space Telescopes (MAST; \url{http://archive.stsci.edu}). The composite spectra, emission-line measurements and derived abundances presented in this work are available from the corresponding author on reasonable request. The \textsc{jwspecfit} code used for the redshift fitting, emission-line fitting and abundance derivations is publicly available at \url{https://github.com/raunaq-rai/jwspecfit}.



\bibliographystyle{mnras}
\bibliography{references} 




\appendix
\section{Physical Properties of all composite spectra}
\label{app:allstacks}

\begin{sidewaystable*}
  \centering
  \vspace*{\fill}
  \footnotesize
  \setlength{\tabcolsep}{3pt}
  \renewcommand{\arraystretch}{1.4}
  \caption{Physical properties of the SFRE- and H$\beta$-EW-binned composite spectra and the full-sample composite ($6 \leq z \leq 10$, $-21<M_{\rm UV}<-19$). Sample columns ($z$, $M_{\rm UV}$, $\log M_\star$) give the median and its 16th--84th percentile spread over the contributing galaxies. Subscripts: Direct = direct-$T_e$, SL = strong-line. $T_e^{\rm high/mid/low}$ are the high- and low-ionisation electron temperatures; $n_e^{\rm low/mid/high}$ the corresponding electron densities.}
  \label{tab:stack_properties_all}
  \resizebox{\textheight}{!}{%
  \begin{tabular}{lcccccccccccccccccc}
    \hline\hline
    Stack & $N$ & Median $z$ & $M_{\rm UV}$ & $\log(M_\star/M_\odot)$ & EW([O\,\textsc{iii}]$+$H$\beta$) & $12+\log(\mathrm{O/H})_{\rm dir}$ & $12+\log(\mathrm{O/H})_{\rm SL}$ & $\log(\mathrm{N/O})$ & $\log(\mathrm{C/O})$ & $\log(\mathrm{Ne/O})$ & $A_V$ & $\log U$ & $T_e^{\rm high}$ (K) & $T_e^{\rm mid}$ (K) & $T_e^{\rm low}$ (K) & $n_e^{\rm low}$ (cm$^{-3}$) & $n_e^{\rm mid}$ (cm$^{-3}$) & $n_e^{\rm high}$ (cm$^{-3}$) \\
    \hline
    \multicolumn{19}{l}{\textit{Full sample}} \\
    \quad Full & 135 & $6.91^{+1.31}_{-0.64}$ & $-19.73^{+0.49}_{-0.67}$ & $8.65^{+0.44}_{-0.39}$ & $788^{+10}_{-10}$ & $7.79^{+0.03}_{-0.03}$ & $7.71^{+0.18}_{-0.15}$ & $-0.57^{+0.09}_{-0.10}$ & $-0.80^{+0.03}_{-0.03}$ & $-0.78^{+0.01}_{-0.01}$ & $0.24^{+0.04}_{-0.04}$ & $-2.97^{+0.01}_{-0.01}$ & $16359^{+350}_{-333}$ & $15278^{+290}_{-277}$ & $14451^{+245}_{-233}$ & $963$ & $27234$ & $59693$ \\
    \multicolumn{19}{l}{\textit{$\Delta t = 3$~Myr}} \\
    \quad Lull & 70 & $6.82^{+1.13}_{-0.55}$ & $-19.82^{+0.46}_{-0.76}$ & $8.74^{+0.42}_{-0.38}$ & $559^{+9}_{-9}$ & $7.99^{+0.05}_{-0.05}$ & $7.89^{+0.25}_{-0.21}$ & $-0.28^{+0.12}_{-0.14}$ & $-0.83^{+0.07}_{-0.06}$ & $-0.74^{+0.02}_{-0.02}$ & $0.18^{+0.06}_{-0.06}$ & $-3.27^{+0.02}_{-0.02}$ & $13957^{+576}_{-525}$ & $13284^{+478}_{-436}$ & $12770^{+403}_{-367}$ & $669$ & $26504$ & $151048$ \\
    \quad Moderate & 37 & $7.04^{+1.36}_{-0.77}$ & $-19.70^{+0.39}_{-0.60}$ & $8.72^{+0.25}_{-0.36}$ & $1022^{+19}_{-18}$ & $7.86^{+0.04}_{-0.04}$ & $7.65^{+0.15}_{-0.13}$ & $-0.69^{+0.15}_{-0.18}$ & $-0.70^{+0.06}_{-0.05}$ & $-0.78^{+0.02}_{-0.02}$ & $0.16^{+0.05}_{-0.05}$ & $-3.03^{+0.02}_{-0.02}$ & $15338^{+500}_{-465}$ & $14431^{+415}_{-386}$ & $13737^{+350}_{-325}$ & $658$ & $11011$ & $120771$ \\
    \quad Bursty & 26 & $7.21^{+0.96}_{-0.95}$ & $-19.58^{+0.50}_{-0.49}$ & $8.25^{+0.54}_{-0.43}$ & $1219^{+29}_{-29}$ & $7.63^{+0.05}_{-0.05}$ & $7.59^{+0.13}_{-0.12}$ & $-0.79^{+0.13}_{-0.16}$ & $-0.79^{+0.06}_{-0.06}$ & $-0.78^{+0.03}_{-0.03}$ & $0.26^{+0.08}_{-0.08}$ & $-3.00^{+0.02}_{-0.02}$ & $18965^{+700}_{-789}$ & $17441^{+581}_{-655}$ & $16276^{+490}_{-552}$ & $676$ & $5720$ & $132129$ \\
    \multicolumn{19}{l}{\textit{$\Delta t = 5$~Myr}} \\
    \quad Lull & 59 & $6.80^{+1.14}_{-0.62}$ & $-19.84^{+0.46}_{-0.85}$ & $8.82^{+0.43}_{-0.39}$ & $555^{+10}_{-10}$ & $8.00^{+0.05}_{-0.05}$ & $7.89^{+0.23}_{-0.21}$ & $-0.45^{+0.14}_{-0.16}$ & $-0.88^{+0.07}_{-0.07}$ & $-0.83^{+0.02}_{-0.02}$ & $0.13^{+0.06}_{-0.06}$ & $-3.26^{+0.02}_{-0.02}$ & $13825^{+576}_{-520}$ & $13175^{+478}_{-432}$ & $12678^{+403}_{-364}$ & $636$ & $19831$ & $150671$ \\
    \quad Moderate & 44 & $6.97^{+1.43}_{-0.50}$ & $-19.72^{+0.50}_{-0.64}$ & $8.63^{+0.29}_{-0.34}$ & $884^{+20}_{-20}$ & $7.77^{+0.05}_{-0.04}$ & $7.61^{+0.15}_{-0.15}$ & $-0.47^{+0.16}_{-0.15}$ & $-0.72^{+0.05}_{-0.05}$ & $-0.69^{+0.02}_{-0.02}$ & $0.20^{+0.05}_{-0.05}$ & $-3.24^{+0.02}_{-0.02}$ & $16037^{+564}_{-553}$ & $15011^{+468}_{-459}$ & $14226^{+395}_{-387}$ & $652$ & $24464$ & $284403$ \\
    \quad Bursty & 30 & $7.24^{+0.97}_{-0.97}$ & $-19.63^{+0.51}_{-0.58}$ & $8.26^{+0.53}_{-0.41}$ & $1227^{+25}_{-25}$ & $7.66^{+0.04}_{-0.04}$ & $7.61^{+0.13}_{-0.13}$ & $-0.83^{+0.14}_{-0.15}$ & $-0.77^{+0.05}_{-0.06}$ & $-0.78^{+0.02}_{-0.03}$ & $0.24^{+0.08}_{-0.08}$ & $-3.04^{+0.02}_{-0.02}$ & $18593^{+741}_{-644}$ & $17132^{+615}_{-534}$ & $16015^{+519}_{-451}$ & $678$ & $4162$ & $159830$ \\
    \multicolumn{19}{l}{\textit{$\Delta t = 10$~Myr}} \\
    \quad Lull & 49 & $6.76^{+1.20}_{-0.58}$ & $-19.80^{+0.48}_{-0.90}$ & $8.89^{+0.46}_{-0.42}$ & $564^{+11}_{-11}$ & $8.08^{+0.06}_{-0.06}$ & $7.90^{+0.22}_{-0.20}$ & $-0.24^{+0.15}_{-0.17}$ & $-0.90^{+0.08}_{-0.08}$ & $-0.85^{+0.03}_{-0.03}$ & $0.23^{+0.08}_{-0.08}$ & $-3.24^{+0.02}_{-0.02}$ & $13210^{+641}_{-595}$ & $12665^{+532}_{-494}$ & $12247^{+449}_{-416}$ & $631$ & $36096$ & $149154$ \\
    \quad Moderate & 55 & $7.26^{+0.98}_{-0.63}$ & $-19.73^{+0.44}_{-0.48}$ & $8.50^{+0.38}_{-0.25}$ & $900^{+18}_{-18}$ & $7.67^{+0.04}_{-0.04}$ & $7.57^{+0.14}_{-0.14}$ & $-0.73^{+0.15}_{-0.19}$ & $-0.83^{+0.05}_{-0.05}$ & $-0.73^{+0.02}_{-0.02}$ & $0.07^{+0.05}_{-0.05}$ & $-2.91^{+0.02}_{-0.02}$ & $17384^{+594}_{-562}$ & $16129^{+493}_{-466}$ & $15169^{+416}_{-393}$ & $1378$ & $14671$ & $65554$ \\
    \quad Bursty & 29 & $6.85^{+1.36}_{-0.66}$ & $-19.68^{+0.52}_{-0.70}$ & $8.43^{+0.40}_{-0.24}$ & $1158^{+27}_{-27}$ & $7.82^{+0.05}_{-0.05}$ & $7.67^{+0.13}_{-0.13}$ & $-0.54^{+0.14}_{-0.17}$ & $-0.78^{+0.06}_{-0.06}$ & $-0.77^{+0.02}_{-0.03}$ & $0.30^{+0.08}_{-0.08}$ & $-3.18^{+0.02}_{-0.02}$ & $16532^{+625}_{-616}$ & $15422^{+519}_{-511}$ & $14573^{+438}_{-431}$ & $640$ & $24391$ & $314219$ \\
    \multicolumn{19}{l}{\textit{$\Delta t = 20$~Myr}} \\
    \quad Lull & 26 & $6.62^{+0.84}_{-0.51}$ & $-19.65^{+0.39}_{-0.54}$ & $8.90^{+0.38}_{-0.42}$ & $532^{+19}_{-19}$ & $7.82^{+0.08}_{-0.08}$ & $7.84^{+0.23}_{-0.20}$ & $-0.33^{+0.16}_{-0.19}$ & $-0.81^{+0.11}_{-0.11}$ & $-0.87^{+0.04}_{-0.04}$ & $0.32^{+0.12}_{-0.12}$ & $-3.12^{+0.03}_{-0.03}$ & $16098^{+1110}_{-938}$ & $15061^{+921}_{-779}$ & $14269^{+777}_{-657}$ & $618$ & $5977$ & $73584$ \\
    \quad Moderate & 88 & $7.03^{+1.27}_{-0.56}$ & $-19.73^{+0.45}_{-0.62}$ & $8.61^{+0.44}_{-0.31}$ & $786^{+12}_{-11}$ & $7.90^{+0.03}_{-0.03}$ & $7.72^{+0.18}_{-0.16}$ & $-0.62^{+0.14}_{-0.14}$ & $-0.79^{+0.04}_{-0.04}$ & $-0.78^{+0.02}_{-0.02}$ & $0.24^{+0.04}_{-0.04}$ & $-3.20^{+0.02}_{-0.01}$ & $15484^{+421}_{-427}$ & $14552^{+349}_{-354}$ & $13839^{+295}_{-299}$ & $7362$ & $35155$ & $181179$ \\
    \quad Bursty & 14 & $6.92^{+1.03}_{-0.65}$ & $-20.06^{+0.97}_{-0.50}$ & $8.49^{+0.32}_{-0.32}$ & $1109^{+44}_{-46}$ & $7.99^{+0.08}_{-0.09}$ & $7.68^{+0.13}_{-0.13}$ & $-0.38^{+0.14}_{-0.13}$ & $-0.77^{+0.08}_{-0.08}$ & $-0.77^{+0.05}_{-0.05}$ & $0.27^{+0.15}_{-0.15}$ & $-3.09^{+0.04}_{-0.04}$ & $14466^{+907}_{-866}$ & $13707^{+753}_{-719}$ & $13126^{+635}_{-606}$ & $647$ & $60223$ & $232814$ \\
    \multicolumn{19}{l}{\textit{H$\beta$ EW}} \\
    \quad Low & 69 & $6.93^{+1.02}_{-0.68}$ & $-19.79^{+0.53}_{-0.77}$ & $8.80^{+0.37}_{-0.43}$ & $445^{+7}_{-7}$ & $8.02^{+0.05}_{-0.05}$ & $7.90^{+0.24}_{-0.22}$ & $-0.13^{+0.14}_{-0.17}$ & $-0.63^{+0.06}_{-0.06}$ & $-0.70^{+0.02}_{-0.02}$ & $0.48^{+0.06}_{-0.06}$ & $-2.85^{+0.02}_{-0.02}$ & $13652^{+544}_{-531}$ & $13031^{+452}_{-441}$ & $12556^{+381}_{-372}$ & $648$ & $40366$ & $25368$ \\
    \quad Mid & 37 & $6.78^{+1.46}_{-0.58}$ & $-19.74^{+0.45}_{-0.65}$ & $8.61^{+0.40}_{-0.28}$ & $1134^{+15}_{-14}$ & $7.83^{+0.03}_{-0.03}$ & $7.70^{+0.15}_{-0.13}$ & $-0.61^{+0.11}_{-0.12}$ & $-0.88^{+0.04}_{-0.04}$ & $-0.81^{+0.02}_{-0.02}$ & $0.00^{+0.04}_{-0.04}$ & $-3.18^{+0.02}_{-0.02}$ & $16266^{+467}_{-499}$ & $15201^{+388}_{-414}$ & $14386^{+327}_{-349}$ & $633$ & $7645$ & $231762$ \\
    \quad High & 25 & $6.93^{+0.97}_{-0.66}$ & $-19.54^{+0.34}_{-0.37}$ & $8.28^{+0.43}_{-0.19}$ & $1629^{+38}_{-37}$ & $7.61^{+0.04}_{-0.04}$ & $7.52^{+0.13}_{-0.13}$ & $-0.84^{+0.17}_{-0.20}$ & $-0.77^{+0.04}_{-0.04}$ & $-0.78^{+0.03}_{-0.03}$ & $0.23^{+0.07}_{-0.07}$ & $-2.76^{+0.03}_{-0.03}$ & $18376^{+692}_{-676}$ & $16952^{+574}_{-561}$ & $15863^{+484}_{-474}$ & $648$ & $10507$ & $47662$ \\
    \hline
  \end{tabular}
  }
  \vspace*{\fill}
\end{sidewaystable*}



\bsp	
\label{lastpage}
\end{document}